\documentclass[12pt]{article}
\usepackage{arxiv}

\RequirePackage{amsthm,amsmath,amsfonts,amssymb,algorithm,algpseudocode,placeins}
\RequirePackage[authoryear]{natbib}
\RequirePackage[colorlinks,citecolor=blue,urlcolor=blue,backref=page,backref=page]{hyperref}
\RequirePackage{graphicx}

\theoremstyle{plain}

\theoremstyle{definition}

\theoremstyle{remark}

\begin{document}

\title{Bayesian Rank-Clustering}

\author{
 Michael Pearce \\
  Department of Mathematics and Statistics\\
  Reed College\\
  Portland, OR \\
  \texttt{michaelpearce@reed.edu} \\
   \And
 Elena A. Erosheva \\
  Department of Statistics, School of Social Work, and the Center for \\Statistics and the Social Sciences\\
  University of Washington\\
  Seattle, WA \\
  \texttt{erosheva@uw.edu} \\
}








\maketitle
\begin{abstract}
Traditional statistical inference on ordinal comparison data results in an overall ranking of objects, e.g., from best to worst, with each object having a unique rank. However, ranks of some objects may not be statistically distinguishable. This could happen due to insufficient data or to the true underlying object qualities being equal. Because uncertainty communication in estimates of overall rankings is notoriously difficult, we take a different approach and allow groups of objects to have equal ranks or be \textit{rank-clustered} in our model. Existing models related to rank-clustering are limited by their inability to handle a variety of ordinal data types, to quantify uncertainty, or by the need to pre-specify the number and size of potential rank-clusters. We solve these limitations through our proposed Bayesian \textit{Rank-Clustered Bradley-Terry-Luce} model. We accommodate rank-clustering via parameter fusion by imposing a novel spike-and-slab prior on object-specific worth parameters in Bradley-Terry-Luce family of distributions for ordinal comparisons. We demonstrate rank-clustering on simulated and real datasets in surveys, elections, and sports analytics.
\end{abstract}

\section{Introduction}\label{introduction}

In a traditional analysis of ordinal data, we assume $I$ judges assess $J$ objects by providing ordinal preferences, $\Pi$. Each judge's ordinal preferences, $\Pi_i$, may be provided in various forms, such as complete rankings, partial rankings, or pairwise comparisons among the available objects or some subset thereof. Standard statistical model families for ranking data such as Mallows \citep{Mallows1957} or Bradley-Terry-Luce \citep{Bradley1952, plackett1975analysis, luce1959individual} derive or estimate the rank of each object whereby each object receives a unique rank. An estimated \textit{overall ranking} then orders all objects from best to worst. Analyses of this kind are used to rank candidates in ranked choice elections,  
\citep{gormley2008mixture,mollica2017bayesian}, sports teams or players in a league using pairwise game outcomes \citep{tutz2015extended,barrientos2023bayesian}, or genes based on ordinal comparisons of genomics data \citep{eliseussen2023rank,vitelli2018probabilistic}.

However, requiring estimated ranks to be unique is not always useful or appropriate. For example, some objects may be equal or indistinguishable in their true quality or ability. Consider an election in which 2 candidates, both of the same political party, are running for an office. If voters express their preferences solely on the basis of party, the candidates are inherently equal in quality.
In another situation, when the number of votes cast is small, estimated ranks assigned to each candidate could exhibit substantial uncertainty, suggesting the candidates are indistinguishable in quality based on the limited number of observed votes. In such situations, allowing for inference to estimate the candidates as having the same rank or be \textit{rank-clustered} may improve interpretability, prediction, and decision-making when analyzing ordinal preferences.

In this paper, we propose a Bayesian framework for ordinal data analysis that estimates an overall ranking of objects with rank-clusters, develop a computationally-efficient Gibbs sampler for estimation, and apply the model to real and simulated data. Specifically, we choose to model observed rank via the Bradley-Terry-Luce (BTL) family of distributions which permits analysis of ordinal preferences in many forms, such as complete rankings, partial rankings, pairwise comparisons, and groupwise comparisons. To induce rank-clusters, we place a novel spike-and-slab fusion prior on the object-specific parameters of BTL distributions. In contrast to existing work related to rank-clustering in the literature, our model requires neither the parameter order nor the number or size of rank-clusters to be known in advance. Instead, these quantities are treated as random variables and estimated simultaneously so that their corresponding uncertainty is naturally reflected in the resulting inferences. 

The rest of the paper is organized as follows. We first review literature related to rank-clustering in Section \ref{background}. Then, we propose the Partition-based Spike-and-Slab Fusion prior and apply it to a BTL model for ordinal data in Section \ref{methods}. We develop a computationally-efficient Gibbs sampler based on reversible jump Markov chain Monte Carlo and demonstrate its accuracy on simulated data in Section \ref{estimation}. To demonstrate a wide variety of methodological benefits of our proposed framework, in Section \ref{applications}, we apply the model to four real datasets: (i) complete rankings of sushi preferences provided by Japanese adults, (ii) partial rankings of 2021 Minneapolis mayoral candidates expressed by voters in a ranked choice election, (iii) complete and partial rankings of policy options from Eurobarometer 34.1, a survey which measures various European attitudes, and (iv) pairwise basketball game outcomes from the 2023-24 season of the National Basketball Association. We conclude with a brief discussion in Section \ref{discussion}.

\section{Background}\label{background}

Before reviewing ordinal comparisons literature, it is helpful to introduce some basic terminology and notation. Rankings are a type of ordinal preferences that denote a relative ordering of objects from best to worst, potentially allowing ties. We use the operator `$\prec$` to denote a strict ordering of two objects; e.g., $A\prec B$ states that object $A$ is strictly preferred to $B$. An object's \textit{rank} is the place it receives in the ranking.\footnote{Although some authors have drawn a distinction between the terms ``ranking'' and ``ordering,'' in this paper we choose to use solely the former in accordance with its popular usage.} Rankings arise in different forms. Given a collection of objects, a ranking is called \textit{complete} when all objects are ranked. In contrast, a ranking is called \textit{partial} when all objects were considered, but only a subset of the most-preferred are ranked (e.g., a top-$5$ ranking). In a partial ranking, we assume that unranked objects are less-preferred than those ranked, but also that the preference order among the unranked objects is unknown. Next, we call a ranking \textit{incomplete} when a judge is asked only to rank a subset of the complete collection of objects. In incomplete rankings, no information can be gleaned regarding objects not considered. For example, if a voter is asked by an election pollster to rank candidates from a single political party, the ranking should provide no information regarding their preferences on candidates from other parties. We call incomplete rankings involving two objects (candidates in the above example) a \textit{pairwise comparison}, and incomplete rankings involving more than two objects a \textit{groupwise comparison}. Rankings may be both partial and incomplete; e.g., it could be a top-3 ranking of mayoral candidates from a specific political party.

Next, we briefly review methods for estimating rank-clusters based on the Bradley-Terry-Luce and Mallows families of ordinal data models in turn. For a more thorough review of these standard model families, see \cite{Marden1996} and \cite{alvo2014statistical}.

\subsection{Methods based on Bradley-Terry-Luce Distributions}

Most work related to rank-clustering utilized the Bradley-Terry-Luce (BTL) family, which comprises the Bradley-Terry and Plackett-Luce distributions and their extensions. The Bradley-Terry model, proposed by \cite{zermelo1929berechnung} and discovered independently by \cite{Bradley1952}, is parameterized by the vector $\omega\in\mathbb{R}_{>0}^J$, in which each $\omega_j$ corresponds to the \textit{worth} of object $j$. Specifically, the Bradley-Terry model specifies the probability that object $i$ will be ranked above object $j$ in pairwise tournament as
\begin{equation}
     P[i \prec j|\omega_i,\omega_j] = \frac{\omega_i}{\omega_i+\omega_j}.
\end{equation}
The Plackett-Luce model \citep{plackett1975analysis} extended the Bradley-Terry to allow for multiple comparisons and has been justified under Luce's Choice Axiom \citep{luce1959individual} and Thurstone's theory of comparative judgement \citep{Thurstone1927,thompson1967use,yellott1977relationship}. In this model, a ranking $\pi=\{1\prec2\prec\dots\prec J\}$ of $J$ objects is assigned probability
\begin{align}
    P[\Pi=\pi|\omega_1,\dots,\omega_J] &= \prod_{j=1}^J \frac{\omega_j}{\sum_{j'=j}^J \omega_{j'}},
\end{align}
where often one sets $\sum_{j} \omega_j=1$ for identifiability. Rankings drawn from the Plackett-Luce model may be interpreted as being created sequentially, where in the first stage an object is selected among all the options, in the second stage an object is selected among all the remaining, and so on. Extensions of distributions in the BTL family have been proposed to capture intricacies in ranked preferences such as order of presentation effects, ties, and covariates \citep{rao1967ties,critchlow1991paired,gormley2010clustering,chapaaan1982exploiting}.

Since BTL distributions have continuous parameters, rank-clusters may be estimated by employing \textit{parameter fusion} or \textit{shrinkage}. \textit{Parameter fusion} is the process of simultaneously estimating parameter values and groups of parameters that should be set equal in value (i.e., ``fusing'' parameters together). \cite{masarotto2012ranking} analyze pairwise comparison data from sports tournaments with parameter fusion techniques under the Bradley-Terry model.
\cite{masarotto2012ranking} estimate an overall ranking of teams with rank-clusters by applying the frequentist \textit{fused lasso} \citep{tibshirani2005sparsity}, in which the absolute difference between every pair of worth parameters is penalized after some data-driven normalization. In this approach, the fused parameters are made equal and thus create a rank-cluster among the corresponding objects. The approach of \cite{masarotto2012ranking} was extended to additional datasets in sports \citep{tutz2015extended} and academic journal rankings \citep{varin2016statistical,vana2016computing}. \cite{jeon2018sparse} argued that shrinkage methods like those proposed by \cite{masarotto2012ranking} and \cite{tutz2015extended} were developed specifically for pairwise comparisons, and thus have inappropriate penalty functions for application to richer kinds of ordinal data like partial or complete rankings. As a result, \cite{jeon2018sparse} proposed a modified regularization penalty that may be applied to partial or complete rankings under the Plackett-Luce model. Relatedly, \cite{hermes2024joint} consider sparse estimation of a Plackett-Luce model with object-level covariates under judge heterogeneity. In their setting, the number of heterogeneous preference groups and the group membership of each judge are assumed fixed and known. To improve efficiency of estimation across groups and predictive performance, they impose a lasso penalty on group-specific covariate coefficients and a simultaneous fused lasso penalty between each pair of group-specific covariate coefficients. We note that the setting studied by \cite{hermes2024joint} is fundamentally different to ours, in that they assume (known) preference heterogeneity among the judges and the presence of object-specific covariates.

Parameter fusion methods for rank-clustering exhibit four distinct disadvantages: First, maximum likelihood estimation of models in the BTL family, even in their simplest forms, often suffers from numerical instability and slow computational speed. As a result, numerous authors have proposed complex algorithms to improve estimation accuracy or speed \citep{Hunter2004, maystre2015fast, turner2020modelling, nguyen2023efficient}. Second, uncertainty quantification is challenging and theoretically tenuous in lasso-based methods \citep{tibshirani1996regression, fan2001variable}. Third, lasso penalty parameters may be difficult to select, requiring data-driven or \textit{ad hoc} techniques \citep{tibshirani1996regression, masarotto2012ranking}. Thus, interpretation of the resulting parameter estimates and associated uncertainty is reliant on the specific choice of penalty parameter. Fourth, prior knowledge on the amount and size of rank-clusters cannot be directly incorporated into the frequentist framework: Although the penalty parameter influences estimation of rank-clusters, the specific meaning of various possible choices is not directly interpretable. 

Many of these disadvantages may be addressed using spike-and-slab priors, a Bayesian approach to variable selection \citep{mitchell1988bayesian, george1997approaches, ishwaran2005spike}. 
Spike-and-slab priors assign weight to both a point-mass at 0 (``spike'') and a continuous density function (``slab''). 
Although the specific formulations of these priors vary, they estimate parameters which are precisely zero in a probabilistic framework that incorporates prior knowledge via interpretable hyperparameters, as opposed to opaque penalty parameters.
However, we are aware of only one variant of this prior class for parameter fusion: \cite{wu2021variable} apply spike-and-slab to differences in successive parameters in a linear regression. 
In their method, the order of parameters from least to greatest in coefficient value must be  known in advance (as in the fused lasso). 
This is not practical in the canonical ordinal data setting because the parameter order is equivalent to the overall ranking, whose estimation is a primary goal. 
Thus, no Bayesian parameter fusions methods exist which may be directly applied to ordinal data analyses with rank-clustering.
Alternatively, one may consider the class of continuous shrinkage priors, which include Bayesian variants of the lasso \citep{park2008bayesian} and fused lasso \citep{casella2010penalized} among others (e.g., \cite{griffin2005alternative,carvalho2010horseshoe,bhattacharya2015dirichlet}). 
However, continuous shrinkage priors do not place positive probability on coefficients (or their differences) being precisely zero.
Thus, parameter fusion must be performed via thresholding the posterior distribution, which is often ad-hoc \citep{porwal2021laplace} and will not be considered in this work.

\subsection{Methods based on Mallows Distributions}

Alternatively, one may consider rank-clustering under the Mallows family of ranking models \citep{Mallows1957}. The Mallows family is parameterized by the overall ranking, $\pi_0$, and a scale parameter $\theta\geq0$ that dictates how likely rankings of a given distance to $\pi_0$ are to be drawn. Specifically, the probability of drawing a ranking $\pi$ from a Mallows$(\pi_0,\theta)$ distribution is
\begin{align}
    P[\Pi=\pi|\pi_0,\theta] &= \frac{e^{-\theta d(\pi,\pi_0)}}{\psi(\theta)}
\end{align}
where $d(\cdot,\cdot)$ is a distance metric and $\psi(\theta)$ is a function which provides an appropriate normalizing constant. Foundational models in the family are defined by their distance metric, with common choices being the Kendall's $\tau$ \citep{Kendall1938} and Spearman's $\rho$ \citep{Spearman1904}. 

To our knowledge, the Clustered Mallows Model proposed by \cite{piancastelli2024clustered} is the only rank-clustering method based on the Mallows model. Their work, proposed concurrently and independently to ours, models \textit{item indifference} (i.e., rank-clusters) by permitting the overall ranking parameter $\pi_0$ to include groups of objects that are tied in rank. The model is estimated in a Bayesian framework from the observed ranking data. However, there are 3 major limitations to their work: First and most importantly, the model requires both the number of rank-clusters and the number of objects per cluster to be pre-specified. Although the authors propose sensible and efficient tools for model selection, the requirement opens the possibility of model misspecification. For example, given 7 objects there are 127 model specifications; given 10 objects there are 1023 model specifications. In addition, pre-specifying the rank-clustering structure removes any uncertainty in the number of rank-clusters and their sizes from the inference task, which we believe to be of key interest in many applications. Second, Bayesian inference of a Clustered Mallows Model is in the class of doubly-intractable problems since the proposed model's normalizing constant is not available in closed form. As a result, exact inference may be computationally slow, or approximation methods may need to be used that require an inexact pseudolikelihood approach. Third, the Mallows model is best suited for ordinal data in the form of complete or partial rankings, meaning the Clustered Mallows Model cannot handle pairwise or groupwise comparisons. As will be shown in Section \ref{methods}, our proposed model avoids all three issues by incorporating parameter fusion in the continuously-parameterized Bradley-Terry-Luce model family.

\section{The Rank-Clustered Bradley-Terry-Luce Model}\label{methods}

In this section, we first develop a novel spike-and-slab prior for parameter fusion based on partitions. Then, we employ the prior in a model for rank-clustering based on the Bradley-Terry-Luce family of ordinal data models.

\subsection{Partition-based Spike-and-Slab Fusion (PSSF) Prior}\label{prior}

Suppose data are drawn exchangeably from a model, $\mathcal{M}$, parameterized by the vector $\omega$. We suppose $\omega$ is of length $J$ and let each $\omega_j\in\Omega$, $\Omega\subseteq\mathbb{R}$. Our goal is to estimate $\omega$ under the belief that some pairs or groups of parameters in $\omega$ may be clustered (i.e., \textit{fused}). We say that two parameters $m,n\in\{1,\dots,J\}$, $m\neq n$, are clustered precisely when $\omega_m=\omega_n$. Clustered parameters may take on any value in their domain, $\Omega$.

Before specifying the prior, we provide some notation on partitions. 
A partition of an object set $\mathcal{J}=\{1,2,\dots,J\}$ is a collection $g=\{C(1),C(2),\dots,C(K)\}$ of $K$ disjoint nonempty subsets (henceforth referred to as ``clusters'') of $\mathcal{J}$ such that their union forms $\mathcal{J}$.
Let $C^{-1}(j)$ represent the cluster that contains object $j\in\mathcal{J}$.
We let $S(k) = \big|\{C(k)\}\big|$ be the size of the subset $C(k)$, and denote by $K$ the number of clusters in $g$.
To emphasize dependence on $g$, we often write $K_g$, $C_g(k)$, etc.
Lastly, we let $\mathcal{G}$ represent the collection of all partitions $g$ of $\mathcal{J}$, and let $\mathcal{G}_k = \{g\in \mathcal{G} : K_g = k\}$.

We are now ready to specify the Partition-based Spike-and-Slab Fusion (PSSF) prior. Under PSSF, $\omega$ is assumed to be generated via the following hierarchical model:
\begin{align}
    G &\sim f_G \nonumber\\
  \nu_k | G=g &\overset{iid}{\sim} f_\nu && k=1,2,\dots,K_g\label{eq:pssfprior}\\
  \omega_j &= \nu_{C^{-1}_g(j)} && j\in\mathcal{J}\nonumber
\end{align}
In Equation \ref{eq:pssfprior}, $f_G(\cdot)$ is a probability mass function on $\mathcal{G}$ and $f_\nu(\cdot)$ is a probability density function on $\Omega$. In words, the prior generates a partition $g$, and then assigns a unique value $\nu_k$ to each cluster $C(k)\in g$. Last, each parameter in $\omega$ is assigned the value of $\nu$ corresponding to its cluster in $g$. 

As an example, suppose $\mathcal{J}=\{1,2,3\}$ and we draw $g=\{C(1),C(2)\}$ such that $C(1)=\{2\}$ and $C(2)=\{1,3\}$, and draw $\nu = [5,10]$. Then, $\omega = [10,5,10]$ because,
\begin{align*}
    \omega_1 &= \nu_{C^{-1}_g(1)} = \nu_{2}=10,\\
    \omega_2 &= \nu_{C^{-1}_g(2)} = \nu_1=5, \text{ and}\\
    \omega_3 &= \nu_{C^{-1}_g(3)} = \nu_2 = 10.
\end{align*}

\subsubsection{Marginal Prior Probabilities}

A useful feature of the PSSF prior is that, regardless of $f_G$, the marginal distribution of each $\omega_j$ follows $f_\nu$. This is because,
\begin{align}
    P[\omega_j] &= \sum_{k=1}^J P[\nu_k|j\in C(k)]P[j\in C(k)]\label{eq:pssf:marginal1}\\
    &= P[\nu_{1}]\sum_{k=1}^J P[j\in C(k)]\label{eq:pssf:marginal2}\\
    &= f_\nu(\cdot).\label{eq:pssf:marginal3}
\end{align}
Equation \ref{eq:pssf:marginal1} holds as there cannot be more than $J$ clusters and each object belongs to precisely one cluster, Equation \ref{eq:pssf:marginal2} holds by the exchangeability of $\nu_{k}$, and Equation \ref{eq:pssf:marginal3} holds since $P[\nu_1] = f_\nu(\cdot)$ by definition and the Law of Total Probability.

\subsubsection{Relationship to Spike-and-Slab}

We have not yet explained the proposed PSSF prior's relationship to the spike-and-slab. It is easiest to understand their connection by considering the joint prior distribution on two arbitrary component parameters, $\omega_m$ and $\omega_n$, such that $m\neq n$. Due to the partitioning structure of parameters in the PSSF prior, there is prior probability associated with a parameter cluster. Thus, their joint prior distribution contains a ``spike'' component along the line $\omega_m=\omega_n$, with density of that line determined by $f_\nu$. Oppositely, given $\omega_m\neq\omega_n$ their joint prior distribution reflects independent draws from $f_\nu$.

\begin{figure}[ht!]
    \centering
    \includegraphics[width=\textwidth]{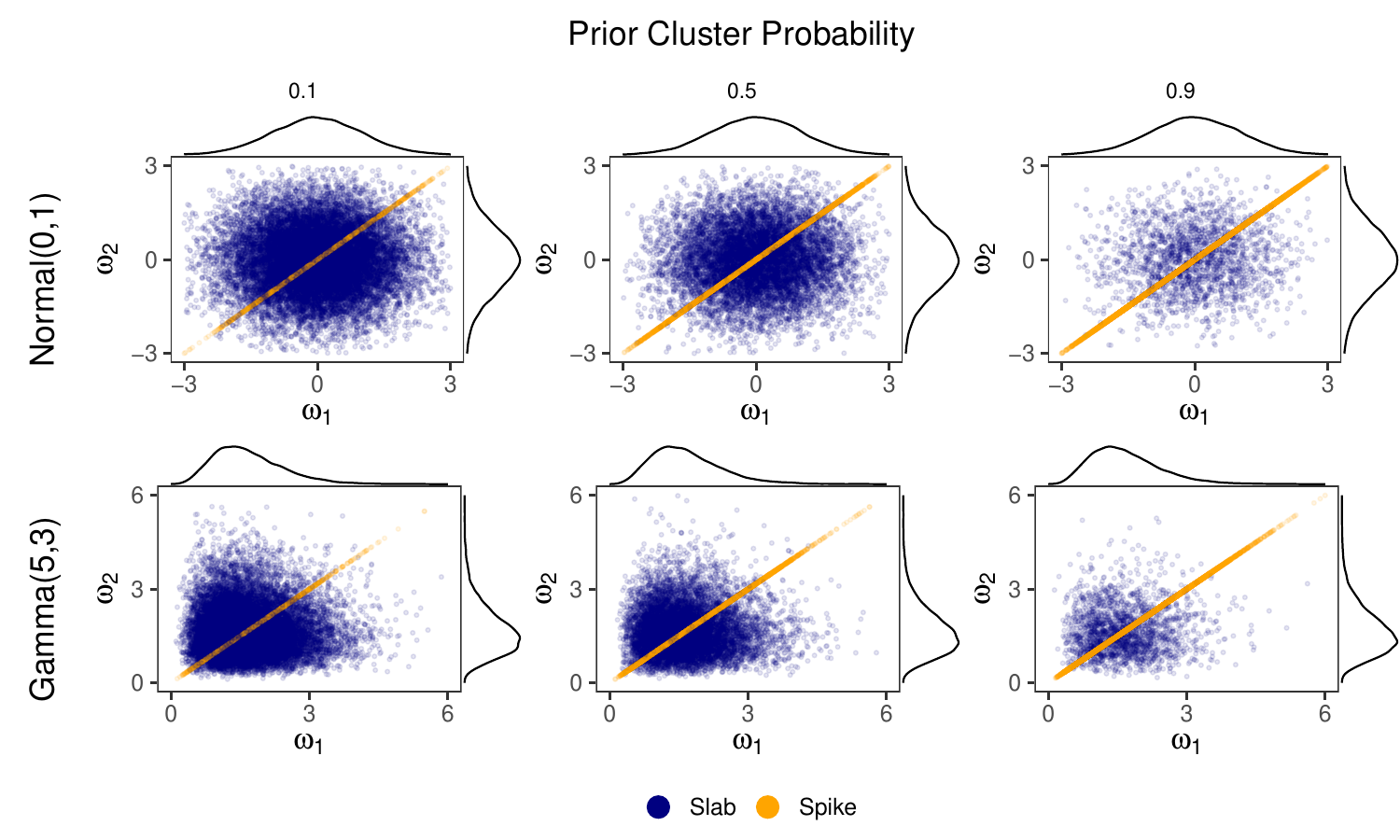}
    \caption[Comparison of PSSF priors under varying specifications.]{Joint distribution of $(\omega_1,\omega_2)$ under the PSSF prior with varying combinations of $f_G$ and $f_\nu$.
    In all cases, $\mathcal{J}=\{1,2\}$, and plots show 20,000 sampled values with marginal density estimates along the axes. Rows correspond to the choice of $f_\nu$ and columns to $f_G$.}
    \label{fig:pssf_example}
\end{figure}

Figure \ref{fig:pssf_example} gives examples of the PSSF prior under varying choices of $f_G$ and $f_\nu$. 
In all panels, we let $\mathcal{J} = \{1,2\}$ and display the joint prior distribution of $(\omega_1,\omega_2)$. In this setting, there are only two unique partitions, $g=\{1,1\}$ and $g=\{1,2\}$. Thus, we specify the prior $f_G$ by stating the so-called ``cluster probability,'' i.e., the probability that $g=\{1,1\}$. Columns correspond to cluster probabilities $0.1, 0.5$, and $0.9$, respectively. Rows correspond to $f_\nu=\text{Normal}(0,1)$ and $\text{Gamma}(5,3)$, respectively. We notice that as the cluster probability increases, so does the density of points in the spike component. Regardless of $f_G$, marginal distributions of each parameter follow $f_\nu$. The marginal relationships seen in Figure \ref{fig:pssf_example} hold identically even as $\mathcal{J}$ grows.

\begin{figure}[ht!]
    \centering
    \includegraphics[width=\textwidth]{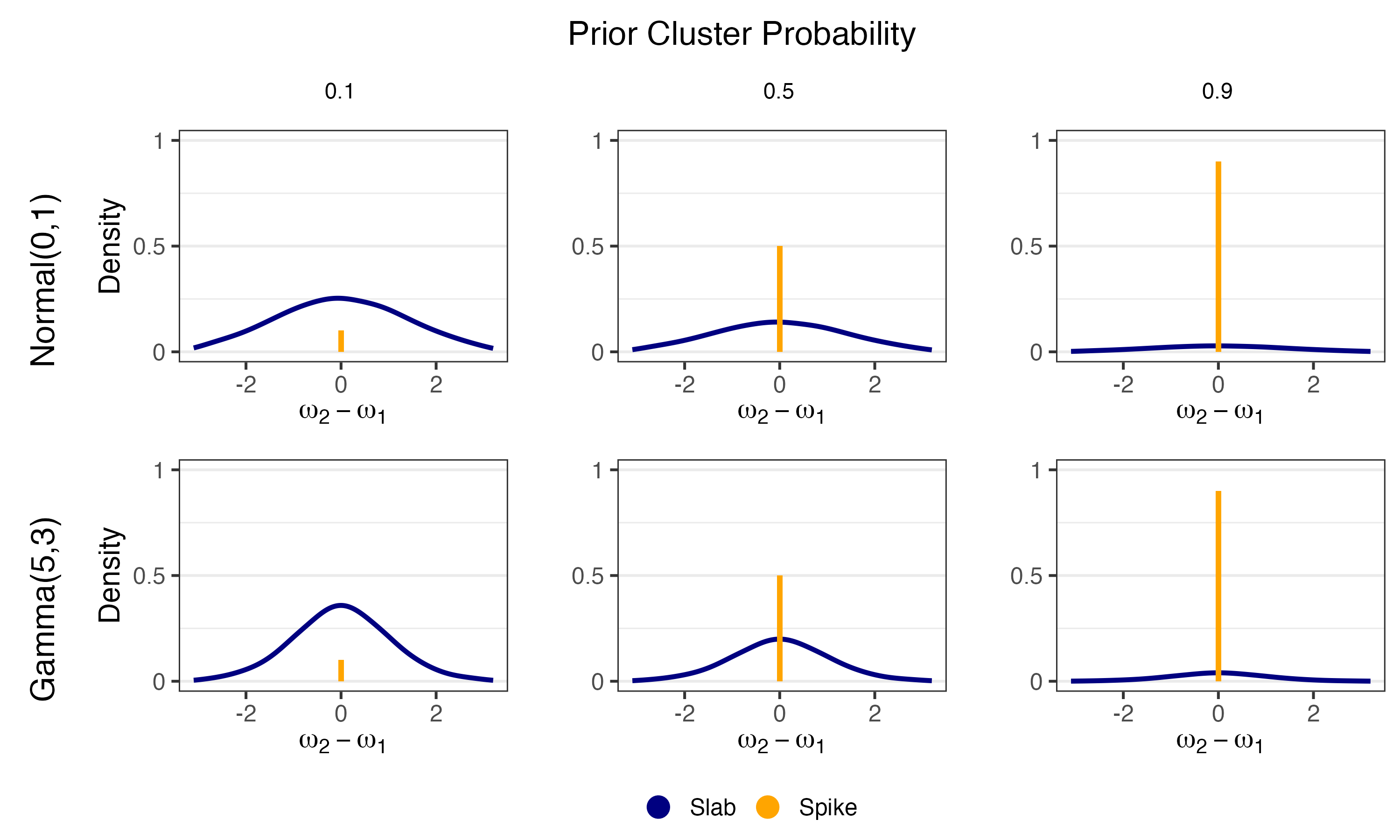}
    \caption[Differences of parameters in PSSF priors under varying specifications.]{Distribution of $\omega_2-\omega_1$ under the PSSF prior with varying combinations of $f_G$ and $f_\nu$.
    In all cases, $\mathcal{J}=\{1,2\}$. Rows correspond to the choice of $f_\nu$ and columns to $f_G$.}
    \label{fig:pdiff}
\end{figure}

Additionally, we display the difference between parameters, $\omega_2-\omega_1$, across different scenarios in Figure \ref{fig:pdiff}. The rows and columns are identical to that from Figure \ref{fig:pssf_example} and make clear the PSSF prior's relationship with the traditional spike-and-slab, which has a spike component at 0 and a background slab density.

\subsection{Rank-Clustered Bradley-Terry-Luce Model}\label{model}

We now introduce the Rank-Clustered Bradley-Terry-Luce model for ordinal data. Let $I$ be the number of judges who assess $J$ objects.  Let $\Pi_i$ represent the ordinal preferences provided by judge $i$, which may be a partial ranking, complete ranking, pairwise comparison, or groupwise comparison. Let $R_i = |\Pi_i|$ and $\mathcal{S}_i$ denote the objects in $\mathcal{J}$ considered by judge $i$, such that $\mathcal{S}_i\subseteq\mathcal{J}$. $R_i$ and $\mathcal{S}_i$ are assumed known.

Under the \textit{Rank-Clustered Bradley-Terry-Luce} (BTL) model, we assume ordinal data is generated via the following Bayesian model:
\begin{align}\label{eq:rankclusterbtl}
    \begin{split}
    \omega &\sim \text{PSSF}(f_G \propto \text{Poisson}(K_g|\lambda),f_\nu = \text{Gamma}(\nu_k|a_\gamma,b_\gamma))\\
    \Pi_i|\omega &\overset{iid}\sim \text{Bradley-Terry-Luce}(\omega|\mathcal{S}_i,R_i)\hspace{4.5cm}i=1,\dots,I
    \end{split}
\end{align}
Rank-Clustered BTL applies the proposed PSSF prior under specific choices of $f_G$ and $f\nu$ to the BTL family of distributions for ordinal data. Note that the data-generating BTL distribution is identifiable up to scalar multiplication of $\omega$. However, the proposed Bayesian model does not suffer from identifiability issues due to the non-uniform prior on $\omega$ \citep{johnson2022bayesian}. We emphasize that the model does not pre-specify the number of clusters, a specific rank-clustering structure, or the order of objects. These are treated as random variables and estimated simultaneously.

\subsubsection{Prior Selection}

We now discuss the selection of priors and hyperparameters. We set $f_G$ according to
\begin{equation}\label{eq:pssf:f_G}
    f_G(g) \propto \text{Poisson}(K_g | \lambda).
\end{equation}
In words, the prior probability of drawing a specific partition $g$ depends only on how many unique clusters, $K_g$, it contains. Thus, every partition with the same $K_g$ has equal prior probability. As a consequence, cluster sizes do not explicitly impact the prior probability of each $g$. Still, there is an implicit connection between cluster size and $K_g$. For example, if $K_g=J$, every cluster must be a singleton. In this setup, one could set $\lambda\approx 1$ to encourage rank-clustering, or $\lambda\approx J$ to discourage rank-clustering.
Next, we set $f_\nu$ according to
\begin{equation}\label{eq:pssf:f_nu}
    f_\nu(\nu_k) = \text{Gamma}(\nu_k|a_\gamma,b_\gamma).
\end{equation}
This Gamma prior has been used in Bayesian estimation of BTL models as it allows for closed-form Gibbs sampling via data augmentation \citep{caron2012efficient,mollica2017bayesian}. The hyperparameters $a_\gamma$ and $b_\gamma$ control the prior distribution on the worth parameters. Since $\omega$ is invariant to multiplicative transformations, $a_\gamma$ and $b_\gamma$ are generally non-influential. Nonetheless, because the ratios between worth parameters could become very large when one object is strongly preferred over another, $(a_\gamma,b_\gamma)$ should be chosen to give some density to values near 0 to allow for such extreme ratios.

\section{Bayesian Estimation}\label{estimation}

In this section, we develop a Gibbs sampler for Bayesian estimation of Rank-Clustered BTL models and provide simulations to demonstrate its performance under varying numbers of observations and rank-clusters.

\subsection{Gibbs Sampler}\label{estimation:gibbs}

Equation \ref{eq:pssfprior} defines $\omega$ by the pair $(\nu,g)$. Thus, to estimate $\omega$, we sample from the joint posterior distribution of $(\nu,g)$. We do so using a reversible jump Markov chain Monte Carlo (RJMCMC) Gibbs sampler that alternates between updating $g$ and $\nu$ via their full conditionals after data augmentation. The sampler is summarized in Algorithm \ref{alg:sparsebtl}.

\begin{algorithm}[ht]
\caption{Gibbs sampler for Rank-Clustered Bradley-Terry-Luce models}\label{alg:sparsebtl}
\begin{enumerate}
    \item Initialize $g^{(0)},\nu^{(0)}$ at random, ensuring that $|\nu^{(0)}| = K_{g^{(0)}}$.
    \item For $t=1,2,\dots,T_1$,
    \begin{enumerate}
        \item Sample $g^{(t)}$ via its full conditional using RJMCMC in order to traverse the space of partitions of varying numbers of clusters.
        \item Sample $\nu^{(t)}$ via its full conditional $T_2$ times, which is possible via closed-form Gibbs sampling with data augmentation.
    \end{enumerate}
\end{enumerate}
\end{algorithm}

Based on our experience fitting Rank-Clustered BTL models to real and simulated data, we recommend initializing $g^{(0)} = \{1,2,\dots,J\}$ (and thus $K_{g^{(0)}}=J$) as it allows rank-clusters to be formed during the estimation process (as opposed to being imposed by the analyst during initialization). For Step 2, $T_1$ should be sufficiently large to allow for convergence of the MCMC chain, although specific choices are context-dependent. Step 2(a) performs RJMCMC on clusters of objects. Since RJMCMC can be slow to converge in high dimensions, it is important to run multiple chains and assess for mixing and convergence \citep{gelman2013bayesian}. Step 2(b) relies on a closed-form Gibbs sampler. We find $T_2\leq 5$ is usually sufficient for posterior sampling.

\subsubsection{Details of Step 2(a)}\label{estimation:gibbs:2a}

We now detail Step 2(a), which proposes a new partition $g'$ based on the current partition $g$. Since $(g,\nu)$ are intricately tied, $\nu$ must simultaneously be updated to an appropriate  $\nu'$. The sampling of discrete partitions is challenging to perform efficiently. In a seminal paper on RJMCMC, \cite{green1995reversible} provided a method for sampling partitions. We adapt that work for the Rank-Clustered BTL model.

Following \cite{green1995reversible}, we only propose $g'$ which are slight modifications of $g$: Precisely, we allow only for `births' splitting one cluster into two, or `deaths' merging two clusters into one. Since all partitions have positive probability, this process is irreducible, as required. There is no need to propose $g'$ that shuffle the partitions but maintain the number of clusters, as these partitions may be obtained by successive birth and death moves.

Births are attempted with probability $b_g = 0.5$.\footnote{One could specify an alternative $b_g\in(0,1)$ or make $b_g$ a function of $K_g$ (as in \cite{green1995reversible}). For simplicity, we fix $b_g=0.5$.} In this case, we select a cluster $k$ at random among those with at least two objects. The cluster is split ``binomially'', meaning that each object is placed independently into one of the ``child'' subgroups, $k_1$ or $k_2$, with equal probability, conditional on each subgroup ultimately containing at least one object. Deaths are attempted with probability $d_g = 1-b_g = 0.5$. In a death, two adjacent clusters are merged at random. Adjacency means that $\not\exists k : \nu_k\in(\nu_{k_1},\nu_{k_2})$.

Births and deaths require updating $\nu$ by increasing or decreasing its dimension by 1, respectively. In a birth, we split a cluster's worth $\nu_k$ into $(\nu'_{k_1},\nu'_{k_2})$ using,
\begin{equation}
    \nu'_{k_1} = u\nu_k, \ \ \ \nu'_{k_2} = u^{-1}\nu_k,
\end{equation}
where $u\sim\text{Unif}(0.5,1.5)$. The corresponding death solves these equations simultaneously:
\begin{equation}
    \nu_k = \sqrt{\nu'_{k_1}\nu'_{k_2}}.
\end{equation}
For reversibility, we automatically reject proposed births where $\nu'_{k_1},\nu'_{k_2}$ are not adjacent.

Per \cite{green1995reversible}, the Metropolis-Hastings probabilities for a birth and death, respectively, are $\min(1,A)$ and $\min(1,A^{-1})$, where
\begin{align}
    A &= \frac{P(\nu',g'|\Pi)}{P(\nu,g|\Pi)}\times \frac{q(\nu,g | \nu',g')}{q(\nu',g' | \nu,g)P(u)}\times \Big|\frac{\partial(\nu'_{k_1},\nu'_{k_2})}{\partial(u,\nu_k)}\Big|,
\end{align}
where $q(\nu',g'|\nu,g)$ is the transition probability of sampling $(\nu',g')$ given current parameter set $(\nu,g)$.
We now calculate each term in $A$. First,

\begin{align}
    \frac{P(\nu',g'|\Pi)}{P(\nu,g|\Pi)} &= \frac{P(\Pi|\nu',g')P[\nu'|g']P[g']}{\sum_{g''}\int_{\nu''}P(\Pi|\nu'',g'')P[\nu''|g'']d\nu''P[g'']}\frac{\sum_{g''}\int_{\nu''}P(\Pi|\nu'',g'')P[\nu''|g'']d\nu''P[g'']}{P(\Pi|\nu,g)P[\nu|g]P[g]}\nonumber\\
    &= \frac{P(\Pi|\nu',g')P[\nu'|g']P[g']}{P(\Pi|\nu,g)P[\nu|g]P[g]}\nonumber\\
    &= \frac{P(\Pi|\nu',g')}{P(\Pi|\nu,g)}\times\frac{\text{Gamma}(\nu'_{k_1}|a_\gamma,b_\gamma)\text{Gamma}(\nu'_{k_2}|a_\gamma,b_\gamma)}{\text{Gamma}(\nu_{k}|a_\gamma,b_\gamma)}\times\frac{P[g']}{P[g]},
\end{align}
where $P(\Pi|\nu,g)$ and $P[g]$ are defined by Equation \ref{eq:rankclusterbtl}.
Second,
\begin{align}\label{eq:Step2a_term2}
\frac{q(\nu,g | \nu',g')}{q(\nu',g' | \nu,g)P(u)} &= \frac{d_{g'}\times\frac{1}{K_{g'}-1}}{\Big(b_g\times\frac{1}{\#\{l:S_{l}(g)\geq2\}}\times\frac{2}{2^{S_g(k)}-2}\Big)\Big(\frac{1}{1.5-0.5}\Big)}\\
&=\frac{d_{g'}\#\{l:S_g(l)\geq2\}(2^{S_g(k)-1}-1)}{b_g(K_{g'}-1)}\nonumber
\end{align}
The numerator in Equation \ref{eq:Step2a_term2} is the death probability, $d_{g'}$, times the probability of selecting a pair of adjacent partitions given $K_{g'}$ total partitions after a split (there are $K_{g'}-1$ such pairs). The denominator is the birth probability, $b_g$, times the probability of selecting a specific cluster $k$ among those with at least two members. This term also includes the probability of dividing the $S_g(k)$ objects in cluster $k$ into two non-empty subsets. There are $(2^{S_g(k)}-2)/2$ such subsets, since there are $2^{S_g(k)}$ total possible partitions, two empty partitions, and two ways to obtain each two-way split.
Third and last,
\begin{align}
\Big|\frac{\partial(\nu'_{k_1},\nu'_{k_2})}{\partial(u,\nu_k)}\Big| &= \Bigg|\begin{bmatrix}
\frac{\partial}{\partial u}\nu_{k_1}' & \frac{\partial}{\partial \nu_k}\nu_{k_1}' \\
\frac{\partial}{\partial u}\nu_{k_2}' & \frac{\partial}{\partial \nu_k}\nu_{k_2}'
\end{bmatrix}\Bigg|= \Bigg|\begin{bmatrix}
\frac{\partial}{\partial u}u\nu_k & \frac{\partial}{\partial \nu_k}u\nu_k \\
\frac{\partial}{\partial u}\nu_k/u & \frac{\partial}{\partial \nu_k}\nu_k/u
\end{bmatrix}\Bigg|=\Bigg|\begin{bmatrix}
\nu_k & u \\
-\nu_k/u^2 & 1/u
\end{bmatrix}\Bigg|\nonumber\\
&=\frac{2\nu_k}{u}.
\end{align}

\subsubsection{Details of Step 2(b)}\label{estimation:gibbs:2b}

To update $\nu$ conditional on a partition $g$ and our data, $\Pi$, we turn to a clever data augmentation trick for Bayesian estimation of Plackett-Luce models as seen in \cite{caron2012efficient} and \cite{mollica2017bayesian}. Here, we adapt their trick to account for the more general BTL family of distributions and rank-clustering. Let $Y = \{Y_{ir}\}$ be a collection of independent random variables, $i=1,\dots,I$ and $r=1,\dots,R_i$, sampled according to
\begin{equation}
    Y_{ir}\sim \text{Exponential}\Big(\sum_{j\in\mathcal{S}_i} \nu_{g^{-1}(j)} - \sum_{s=0}^{r-1}\nu_{g^{-1}(\pi_i(s))}\Big).
\end{equation}
The exponential rates are precisely the denominator terms from BTL densities that are burdensome to calculate. The full conditional posterior probability $P[\nu|Y,\Pi,g]$ is then,
\begin{align}
    P[\nu|Y,\Pi,g] \propto& P[Y|\Pi,g,\nu]P[\Pi|g,\nu]P[g|\nu]P[\nu]\nonumber\\
    \propto& P[Y|\Pi,g,\nu]P[\Pi|g,\nu]P[\nu]\nonumber\\
    =& \prod_{i=1}^I \prod_{r=1}^{R_i} \Big(\sum_{j\in\mathcal{S}_i} \nu_{g^{-1}(j)} - \sum_{s=0}^{r-1}\nu_{g^{-1}(\pi_i(s))}\Big)e^{-y_{ir}\big(\sum_{j\in\mathcal{S}_i} \nu_{g^{-1}(j)} - \sum_{s=0}^{r-1}\nu_{g^{-1}(\pi_i(s))}\big)}\times \nonumber\\
    & \prod_{i=1}^I \prod_{r=1}^{R_i} \frac{\nu_{g^{-1}(\pi_i(r))}}{\sum_{j\in\mathcal{S}_i} \nu_{g^{-1}(j)} - \sum_{s=0}^{r-1}\nu_{g^{-1}(\pi_i(s))}}\times \prod_{k=1}^K \nu_k^{a_\gamma-1}e^{-b_\gamma\nu_k}\nonumber\\
    =& \prod_{i=1}^I \prod_{r=1}^{R_i} \nu_{g^{-1}(\pi_i(r))}e^{-y_{ir}\big(\sum_{j\in\mathcal{S}_i} \nu_{g^{-1}(j)} - \sum_{s=0}^{r-1}\nu_{g^{-1}(\pi_i(s))}\big)}\times\prod_{k=1}^K \nu_k^{a_\gamma-1}e^{-b_\gamma\nu_k}
\end{align}
Given these cancellations, we notice a closed-form expression for the posterior:
\begin{align}
    P[\nu|Y,\Pi,g] \propto& \prod_{i=1}^I\prod_{k=1}^K \nu_k^{c_{ki}}e^{-\nu_k\sum_{r=1}^{R_i}y_{ir}\delta_{irk}}\times\prod_{k=1}^K \nu_k^{a_\gamma-1}e^{-b_\gamma\nu_k}\nonumber\\
    =& \prod_{k=1}^K \nu_k^{a_\gamma+\sum\limits_{i=1}^I c_{ki}-1}e^{-\nu_k(b_\gamma + \sum\limits_{i=1}^I\sum\limits_{r=1}^{R_i}y_{ir}\delta_{irk})}\nonumber\\
    \propto& \prod_{k=1}^K \text{Gamma}\Big(\nu_k \ \Big| \ a_\gamma+\sum\limits_{i=1}^I c_{ki}, b_\gamma + \sum\limits_{i=1}^I\sum\limits_{r=1}^{R_i}y_{ir}\delta_{irk}\Big)
\end{align}
where
\begin{align}
     c_{ki} &= \big|\big\{j : j\in\pi_i, g^{-1}(j) = k\big\}\big|\\
     \delta_{irk} &= \big|\big\{j : j\in\mathcal{S}_i, j\not\in \{\pi_i(1),\dots,\pi_i(r-1)\}, g^{-1}(j) = k\big\}\big|.
\end{align}
Thus, we can sample $\nu$ from a closed-form Gamma distribution after augmentation of the conditioning data $\Pi$ and random variable $g$ with $Y$.

Now that we have developed an efficient estimation algorithm for Rank-Clustered BTL models, we turn to a numerical simulation to demonstrate estimation accuracy under different rank-clustering regimes.

\subsection{Numerical Simulation}\label{estimation:simulation}

We now demonstrate accurate estimation of worth parameters and rank-clusters via a Rank-Clustered BTL model in a numerical simulation. We assume there are $J=8$ objects which form $K$=1, 2, 4, or 8 rank-clusters. When $K=J=8$, every object is independent; there are only singleton rank-clusters. In the true worth parameter vector, $\omega_0$, rank-clustered objects have identical values and successive rank-clusters are separated in value by a factor of 4 (see Table \ref{tab:pssf:omega0} for specific values). Fourfold increases induce strong but not absolute separation between objects: For demonstration, in a pairwise tournament between an object with $\omega_1=1$ and $\omega_2=4$, the probability of selecting object 2 is,
$$P[2\prec 1|\omega_1=1,\omega_2=4] = \frac{\omega_2}{\omega_1+\omega_2}=\frac{4}{4+1} = 0.8.$$
We also vary the Poisson hyperparameter on the number of rank-clusters, $\lambda\in\{0.1,2,4,8\}$, which encourages rank-clustering to different extents and allows us to measure robustness of results when $\lambda$ is somewhat misspecified. Finally, we vary the number of judges $I\in\{50,200,800\}$. For each combination of $I$, $\omega_0$, and $\lambda$, we generate 20 independent datasets and fit a Rank-Clustered Bradley-Terry-Luce distribution to each, under hyperparameters $a_\gamma=5$ and $b_\gamma=3$. We set $T_1=5{,}000$ and $T_2=2$ to obtain 10,000 posterior samples in each MCMC chain and remove the first half as burn-in. For identifiability, posterior estimates of $\omega_0$ are normalized \textit{post-hoc} such that $\sum_j \omega_{0j} = 1$.
\begin{table}[ht]
    \centering
    \begin{tabular}{ll}
         \text{Setting:} & $\omega_0$ \\
         \hline
         \text{$K=1$} & $\{4^0,4^0,4^0,4^0,4^0,4^0,4^0,4^0\}$ \\
         \text{$K=2$} & $\{4^0,4^0,4^0,4^0,4^1,4^1,4^1,4^1\}$ \\
         \text{$K=4$} & $\{4^0,4^0,4^1,4^1,4^2,4^2,4^3,4^3\}$ \\
         \text{$K=8$} & $\{4^0,4^1,4^2,4^3,4^4,4^5,4^6,4^7\}$ \\
    \end{tabular}
    \caption{Simulation settings for $\omega_0$ under varying numbers of true rank-clusters, $K$.}
    \label{tab:pssf:omega0}
\end{table}

\begin{figure}[ht!]
    \centering
    \includegraphics[width=\textwidth]{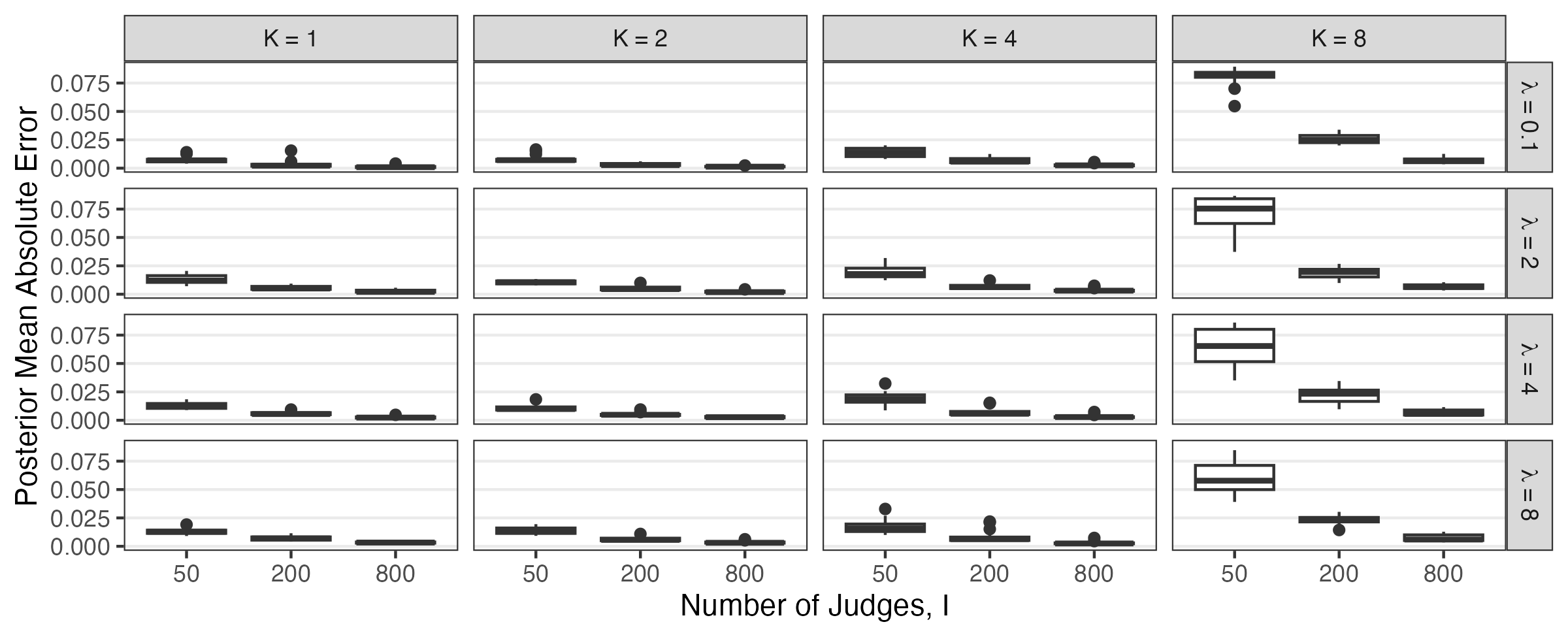}
    \caption{Boxplots of posterior mean absolute error for $\omega_0$ across combinations of the number of judges $I$, true number of rank-clusters $K$, and hyperparameter $\lambda$. Errors are calculated after normalization of posterior samples such that $\sum_{j}\omega_{0j}=1$.}
    \label{fig:pssf:sim1}
\end{figure}

We first examine the accuracy of estimation for $\omega_0$ across simulation settings. Figure \ref{fig:pssf:sim1} displays boxplots of mean absolute error (MAE) for $\omega_0$ by number of judges $I$, true number of rank-clusters $K$, and the choice of hyperparameter $\lambda$. In general, estimation is quite accurate. We see that for any specific combination of $K$ and $\lambda$, MAE decreases as $I$ increases. Estimation error is higher when $K$ is large and $I$ is small, most likely the result of error estimating a complex rank-clustering structure.

\begin{figure}[ht!]
    \centering
    \includegraphics[width=\textwidth]{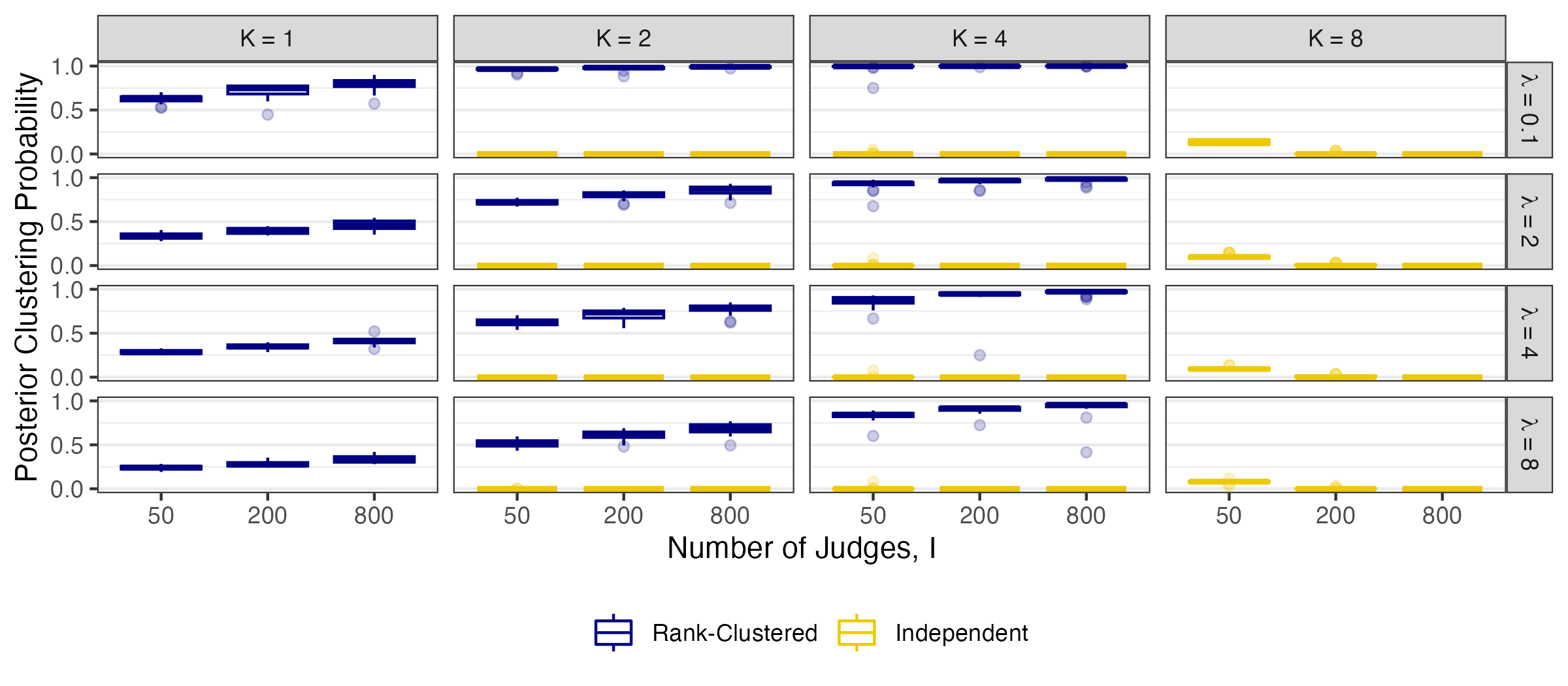}
    \caption{Boxplots of the mean posterior probability of rank-clustering  object pairs which are truly rank-clustered (navy) or independent (gold), across combinations of $I$, $K$, and $\lambda$.}
    \label{fig:pssf:sim2}
\end{figure}

Figure \ref{fig:pssf:sim2} displays the mean posterior probability of rank-clustering across object pairs which are truly rank-clustered (navy) or independent (gold) in $\omega_0$. 
Results are further separated by the number of judges, $I$, true number of clusters, $K$, and hyperparameter $\lambda$. For rank-clustered pairs, accuracy of recovery is generally high and increases with the number of judges, $I$. Accuracy is best when hyperparameter $\lambda\approx K$, which occurs when prior belief regarding the number of rank-clusters is approximately correct. If there is limited prior knowledge on the number of rank-clusters, we suggest specifying a vague hyperparameter setting such as $\lambda = \frac{J}{2}$ and assessing sensitivity of results to various choices of $\lambda$. The posterior probability of rank-clustering independent object pairs is near 0 in all simulations, indicating excellent recovery accuracy of objects with distinct worth parameters.

The numerical simulations in this section indicate that the proposed Rank-Clustered BTL model is able to accurately estimate the relative worth of objects in a collection, including in the presence of rank-clustering. Estimation error decreases to 0 as the number of observations increases. Overall, the model correctly identifies rank-clustered and independent object pairs. 

\section{Applications}\label{applications}

In this section, we apply the Rank-Clustered BTL model to four real datasets involving ordinal comparisons. These four applications were chosen to highlight the applicability of our method to various ordinal data types and domain areas and illustrate methodological values of our approach which are summarized in Table \ref{tab:applications}. The data sets are comprised of sushi preferences of Japanese adults \citep{kamishima2003nantonac}, policy preferences of respondents from Great Britain in a Eurobarometer survey \citep{reif1993euro}, ranked-choice votes in a Minneapolis mayoral election \citep{Minneapolis2021}, and pairwise game outcomes among teams in the United States National Basketball Association \citep{nba2023_24}. 
\begin{table}[ht]
    \centering
    \begin{tabular}{p{.04\linewidth} | p{0.16\linewidth} | p{0.21\linewidth} | p{0.46\linewidth}}
          & \text{Setting} & \text{Data Type} &\text{Methodological Value} \\
         \hline
         5.1 & Sushi preferences in Tohoku & Complete rankings of 10 sushi types & Rank-clusters sushi types by preferences. Comparing inferred overall ranking with that of the Clustered Mallows Model.\\
         5.2 & Minneapolis mayoral election votes & Top-3 partial rankings of 17 candidates & Interpretable overall ranking captures the winner's mandate in ranked-choice elections. Comparing inferred overall ranking with those from a standard BTL model and two election procedures.\\
         5.3 & Eurobarometer survey policy preferences & Partial rankings of 7 policy options & Inferred overall ranking permits identification of similarly preferred options to aid policymakers.\\
         5.4 & Basketball game outcomes & Pairwise comparisons (game winners) among 30 teams & Inferred overall order captures similarly-performing teams. Setting with limited information and low signal.\\
    \end{tabular}
    \caption{Summary of applications by subsection.}
    \label{tab:applications}
\end{table}

\subsection{Sushi Preferences in Tohoku}

\begin{figure}[ht!]
    \centering
    \includegraphics[width=\textwidth]{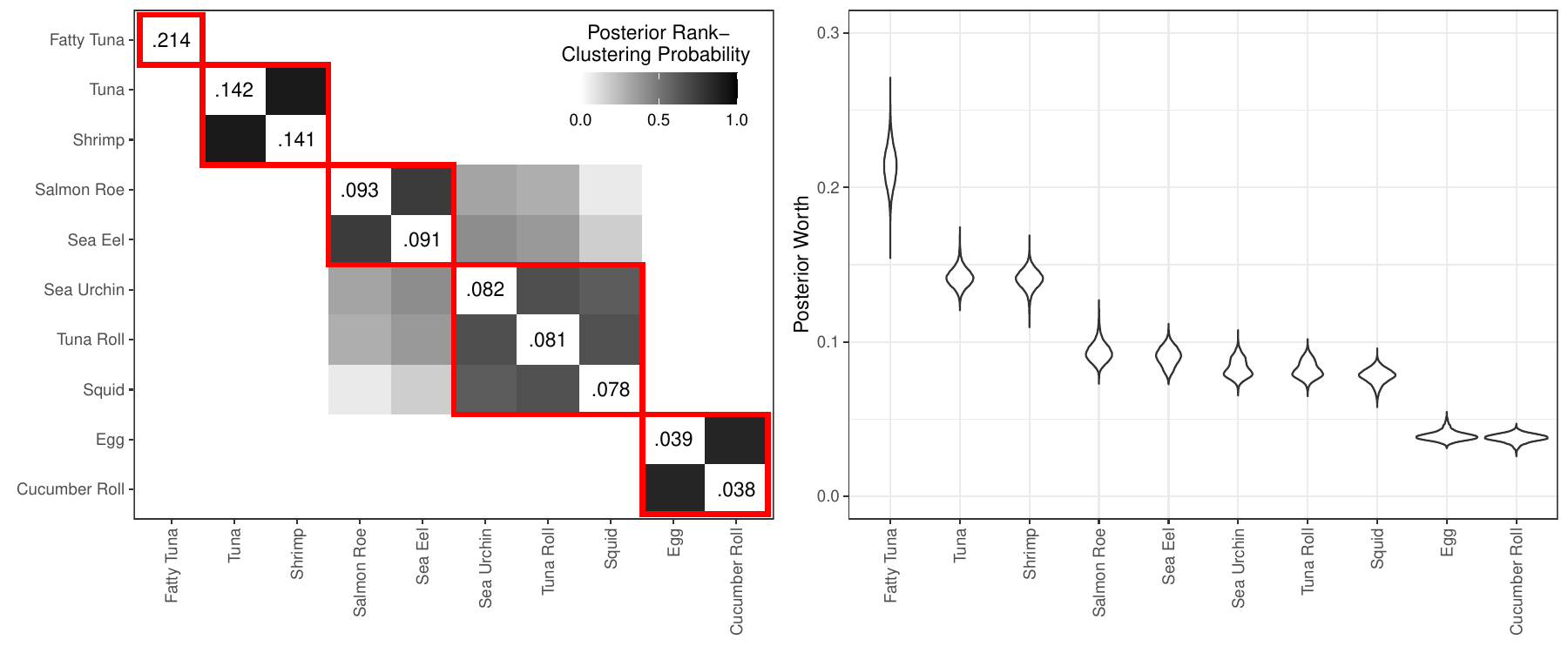}
    \caption{Primary results from Rank-Clustered BTL analysis of Tohoku sushi data. \textit{Left:} Posterior rank-clustering probabilities. Main diagonal displays posterior median estimate of worth parameter after normalization. Red squares indicate maximum \textit{a posteriori} rank-clusters. \textit{Right:} Posterior distributions of sushi-specific worth parameters.}
    \label{fig:Sushi_1}
\end{figure}

We first study complete preference rankings of 10 sushi types from a benchmarking dataset by \cite{kamishima2003nantonac}. To allow our results to be comparable with an analysis of the sushi data by \cite{piancastelli2024clustered}, we analyze the preferences of survey respondents who lived in Japan's Tohoku region until at least 15 years of age. There were 280 such respondents.
We fit a Rank-Clustered BTL distribution to the data with $a_\gamma=5$, $b_\gamma=3$, and $\lambda = 2$ to encourage rank-clustering but permit a wide variety of outcomes. Figure \ref{fig:Sushi_1} displays posterior rank-clustering probabilities (left) and parameter posteriors (right). Additional results and convergence diagnostics are provided in the supplementary materials.

Sushi types are ordered according to posterior median worth. Based on the left panel in Figure \ref{fig:Sushi_1}, fatty tuna appears to be strictly most preferred in this population, followed by tuna and shrimp rank-clustered in second place. Salmon roe and sea eel exhibit high posterior probability of rank-clustering, as do sea urchin, tuna roll, and squid; these two groups may themselves be rank-clustered. Egg and cucumber roll are rank-clustered in last place. Our results demonstrate the proposed model's ability to rank-cluster objects with uncertainty under complete rankings in survey data.

We compare our results to those found by \cite{piancastelli2024clustered} in a Clustered Mallows Model (CMM). They estimate the following ranking: fatty tuna $\prec$ tuna $\prec$ shrimp $\prec$ \{salmon roe, sea urchin\} $\prec$ \{sea eel, tuna roll, squid\} $\prec$ \{egg, cucumber roll\}. Our results are, unsurprisingly, similar, but differ in illuminating ways. Tuna and shrimp are rank-clustered in our model. The rank-clusters \{salmon roe, sea eel\} and \{sea urchin, tuna roll, squid\} swap the rank of sea eel and sea urchin. These two rank-clusters exhibit some posterior probability of rank-clustering themselves. These differences showcase how the model pre-specification required by CMM limits the flexibility of results and may not fully show what the data has to offer or fully account for uncertainty in the estimated ranks and rank-clusters. The Rank-Clustered BTL model
requires no pre-specification and permits complex posterior summaries of rank-clustering, including uncertainty in the number of rank-clusters and their respective sizes.

\subsection{2021 Minneapolis Mayoral Election}

\begin{figure}[ht!]
    \centering
    \includegraphics[width=\textwidth]{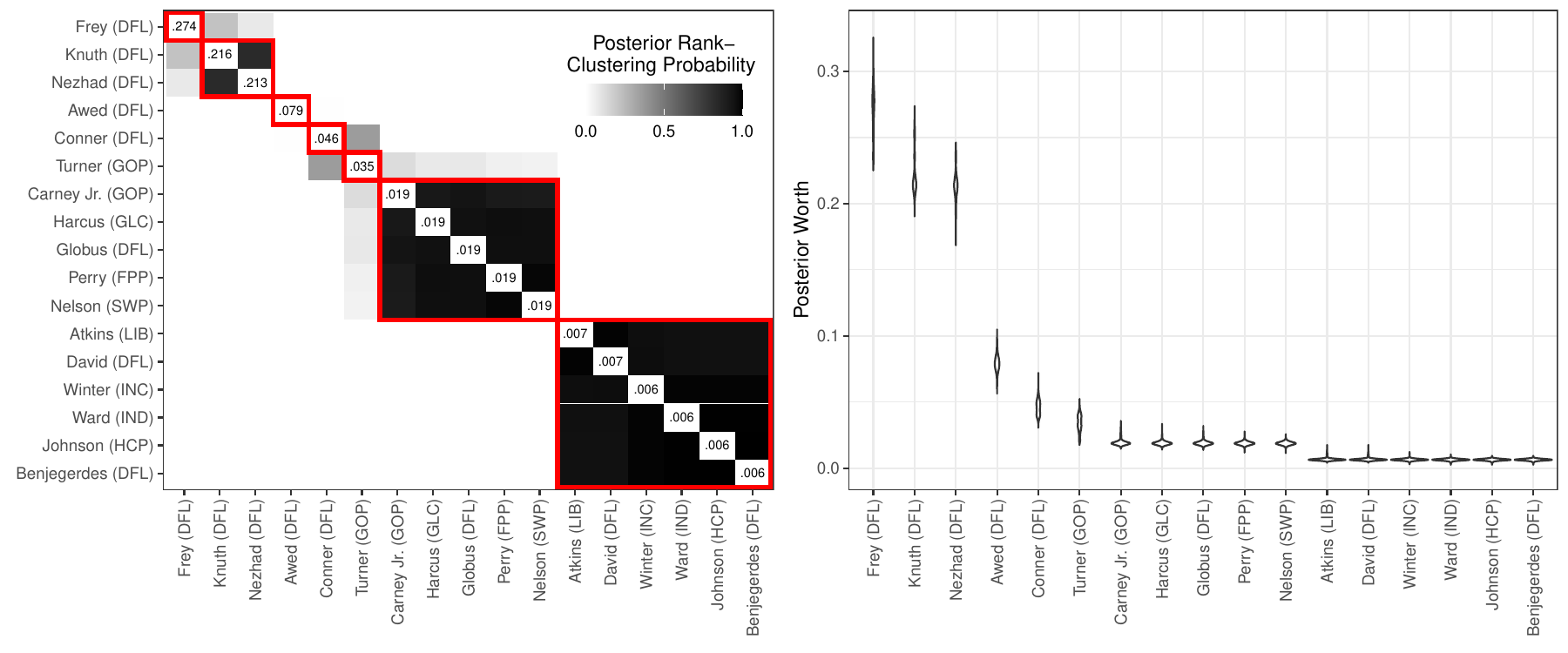}
    \caption{Primary results from Rank-Clustered BTL analysis of mayoral votes. \textit{Left:} Posterior rank-clustering probabilities. Main diagonal displays posterior median estimate of worth parameter after normalization. Red squares indicate maximum \textit{a posteriori} rank-clusters. \textit{Right:} Posterior distributions of candidate-specific worth parameters.}
    \label{minneapolis_res}
\end{figure}

Our second example analyzes real rank-choice votes from the 2021 mayoral election in Minneapolis, Minnesota \citep{Minneapolis2021}. This election included 17 candidates (excluding write-ins and one who received no votes) and asked voters to rank their top-3 choices, in order. 
A total of 145,337 votes were cast in this election. To mimic exit polling data, we randomly sample 1000 valid votes for analysis, which we treat as a random sample of preferences from the population of Minneapolis voters.
We want to estimate the overall preferences of Minneapolis voters regarding mayoral candidates and learn which candidates, if any, are rank-clustered at the population level.
Clustering candidates may be of interest to political scientists or local political organizations for the purpose of understanding voter preferences \citep{gunther2003species, dimock2014beyond}. For example, if the winner of the election is deemed to be rank-clustered with other candidate(s), their mandate may be considered weak. Conversely, if the winner is a singleton first-place rank-cluster---clearly ranked above all other candidates---their mandate may be considered strong.
We fit a Rank-Clustered BTL to the data with $a_\gamma=5$, $b_\gamma=3$, and $\lambda = 2$ to encourage few rank-clusters. Figure \ref{minneapolis_res} displays posterior rank-clustering probabilities (left) and parameter posteriors (right). Additional results and convergence diagnostics are provided in the supplementary materials.

In Figure \ref{minneapolis_res}, candidates are ordered by their posterior median estimate of worth. Cluster 1 consists of Jacob Frey, the winner and incumbent. We note that Frey is not rank-clustered with other candidates with high posterior probability, suggesting a relatively strong mandate. 
Cluster 2 consists of Kate Knuth and Sheila Nezhad, both female, non-incumbent DFL candidates. 
Last, Cluster 7 consists of 6 candidates with minimal support.

\begin{figure}[t!]
    \centering
    \includegraphics[width=\textwidth]{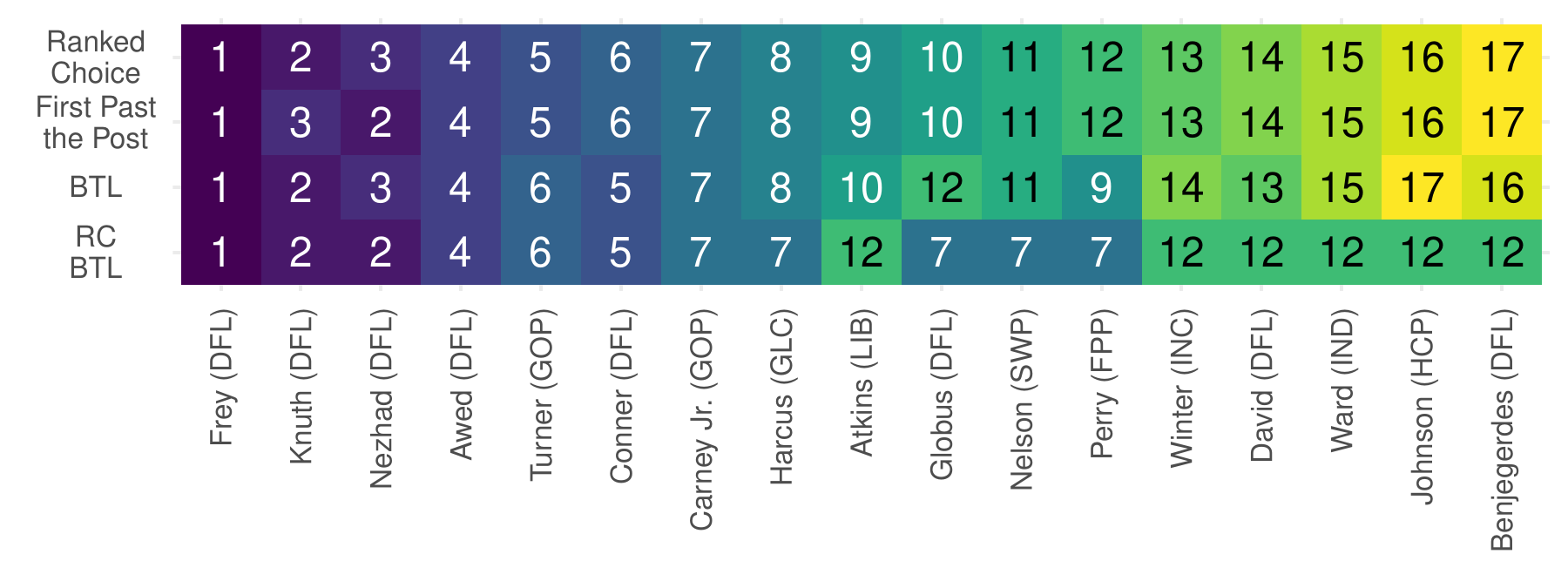}
    \caption[Comparison of estimated candidate ranks among four methods on the 2021 Minneapolis mayoral election data.]{Comparison of estimated rank for each candidate across four aggregation methods: Ranked Choice, First-Past-the-Post (FPP), BTL, and Rank-Clustered BTL (RC BTL). Candidates are ordered by their rank in the actual ranked choice election.}
    \label{fig:votes_compare}
\end{figure}

Figure \ref{fig:votes_compare} compares point estimates of rank for each candidate across four methods.
The first and second rows display assigned ranks from ranked choice and ``first-past-the-post'' (FPP) election procedures, respectively.
We calculate FPP ranks by ordering candidates by the number of first place votes he/she received (ignoring all second and third place votes).\footnote{If the actual election had utilized FPP tabulation, results may have been different based on the differing voter strategies encouraged by ranked choice and FPP elections.} The third and fourth rows display maximum \textit{a posteriori} ranks from a standard Bayesian BTL and our Rank-Clustered BTL, respectively. Frey wins the election in all methods. 
The BTL and Rank-Clustered BTL models roughly reflect the deterministic algorithms, although we notice some swaps in candidate ranks which may be attributed to differences between first place and second or third place votes. For example, Conner received fewer first place votes than Turner, but far more second and third place votes (see supplementary materials for vote totals). As a result, deterministic algorithms rank Turner above Conner, while the BTL model takes into account the additional preference information and ranks Conner above Turner. 
In summary, the overall ordering estimated by the Rank-Clustered BTL differs from a standard BTL model and two deterministic election procedures. Furthermore, our model confirms that Frey is strictly preferred over the remaining candidates by voters. 

\FloatBarrier

\subsection{Eurobarometer 34.1 Survey Data}

We analyze data from the Eurobarometer 34.1 survey \citep{reif1993euro}, which included the following question:

\begin{quote}
    \underline{Question 28:} There are various actions that could be taken to eliminate the drugs problem. In your opinion, what is the first priority? And the next most urgent? (Ask respondent to rank all 7, with 1 as the most urgent.)
\begin{enumerate}
    \item Information campaigns about the dangers of drugs.
    \item Hunting down drug pushers and distributors.
    \item Legal penalty for drug taking.
    \item Looking after and treating drug addicts and rehabilitating them.
    \item Funding research into drug substitutes, and into the treatment of drug addiction.
    \item Fighting the social causes of drug addiction.
    \item Reinforcing the control or distribution and usage of addictive medicines.
\end{enumerate}
\end{quote}
We subset the data to respondents from Great Britain to avoid heterogeneity and non-proportional sampling among respondents from different European countries. There were 1005 valid responses among this group (out of 1031 total surveyed), of which 970 were complete rankings and the rest ranked between 1 and 5 items (a top-6 ranking is inherently equivalent to a complete ranking since all survey options were presented). We seek to identify a population-level ordering of the priorities that accounts for potential equality or indistinguishability among the options based on the survey data. These data were previously studied by \cite{wang2017variational} with a mixed-membership model to learn about heterogeneity of opinions among survey respondents. Our analysis, although a simplification of the diverse population's heterogeneous preferences, provides a simpler interpretation to policy-makers interested in understanding rank-ordering of policy preferences. 

\begin{figure}[ht!]
    \centering
    \includegraphics[width=\textwidth]{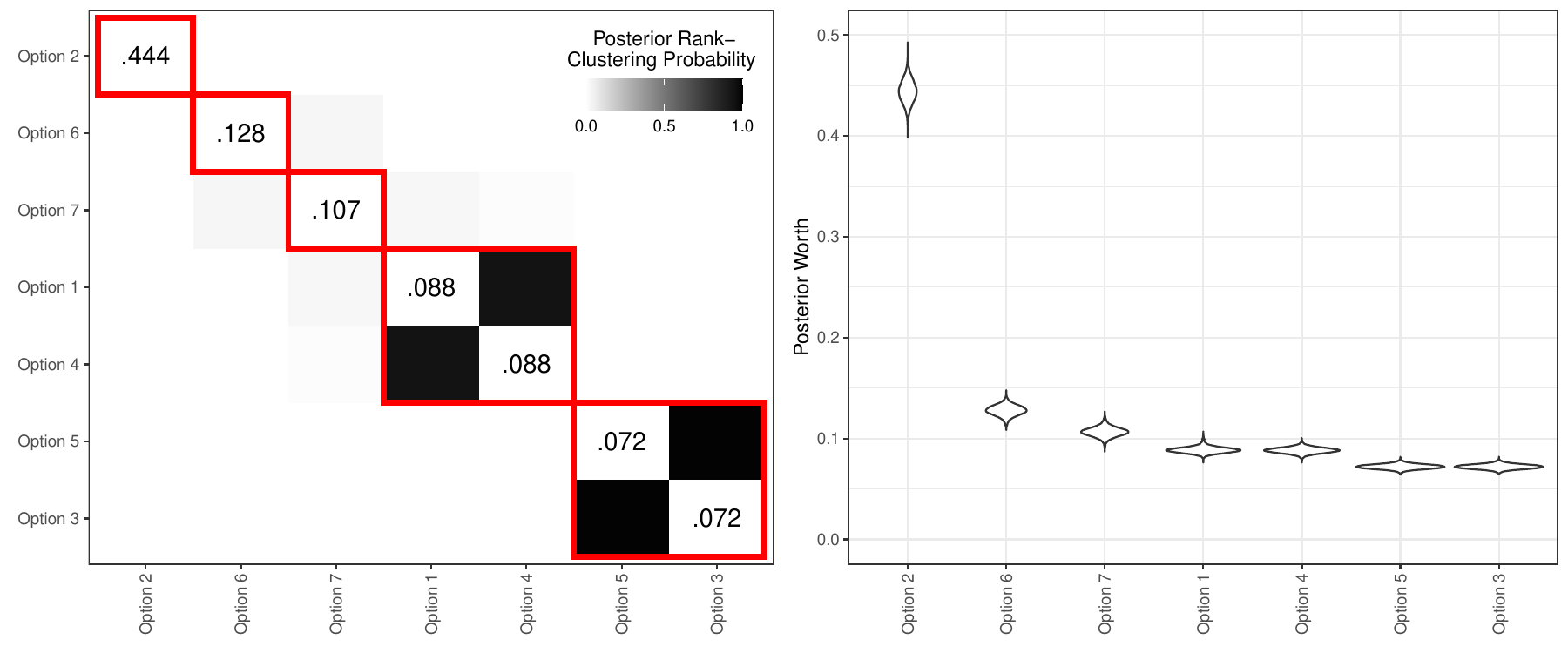}
    \caption{Primary results from Rank-Clustered BTL analysis of Eurobarometer 34.1 data. \textit{Left:} Posterior rank-clustering probabilities. Main diagonal displays posterior median estimate of worth parameter after normalization. Red squares indicate maximum \textit{a posteriori} rank-clusters. \textit{Right:} Posterior distributions of policy-specific worth parameters.}
    \label{fig:EB_1}
\end{figure}

We fit a Rank-Clustered Bradley-Terry-Luce model to the data with $a_\gamma=5$, $b_\gamma=3$, and $\lambda=2$ to encourage rank-clustering. Figure \ref{fig:EB_1} displays posterior rank-clustering probabilities (left) and parameter posteriors (right). Additional results and convergence diagnostics provided in the supplementary materials.

Policy option 2 (\textit{information campaigns}) is strictly preferred to the rest among the population of survey respondents from Great Britain, whereas options 5 (\textit{funding research}) and 3 (\textit{legal penalty}) are rank-clustered last. The results indicates to policymakers that respondents in Great Britain strongly prioritize Option 2 in comparison to the rest, while pairs of Options 1 and 4 and Options 3 and 5, are, respectively, indistinguishable within each pair, with 1 and 4 being strongly preferred to 3 and 5. By rank-clustering similarly-preferred options, interpretation of constituent preferences is simplified for policymakers.

\subsection{2023-24 National Basketball Association Game Outcomes}

Last, we analyze outcomes of 1$,$230 games from the 2023-24 season of the National Basketball Association (NBA) of the United States of America \citep{nba2023_24}. In this season, 30 teams each played 82 games, including between 2 and 5 games against every other team. We seek to estimate an overall ranking of teams that allows for potential equality in ranking. 

\begin{figure}[ht!]
    \centering
    \includegraphics[width=\textwidth]{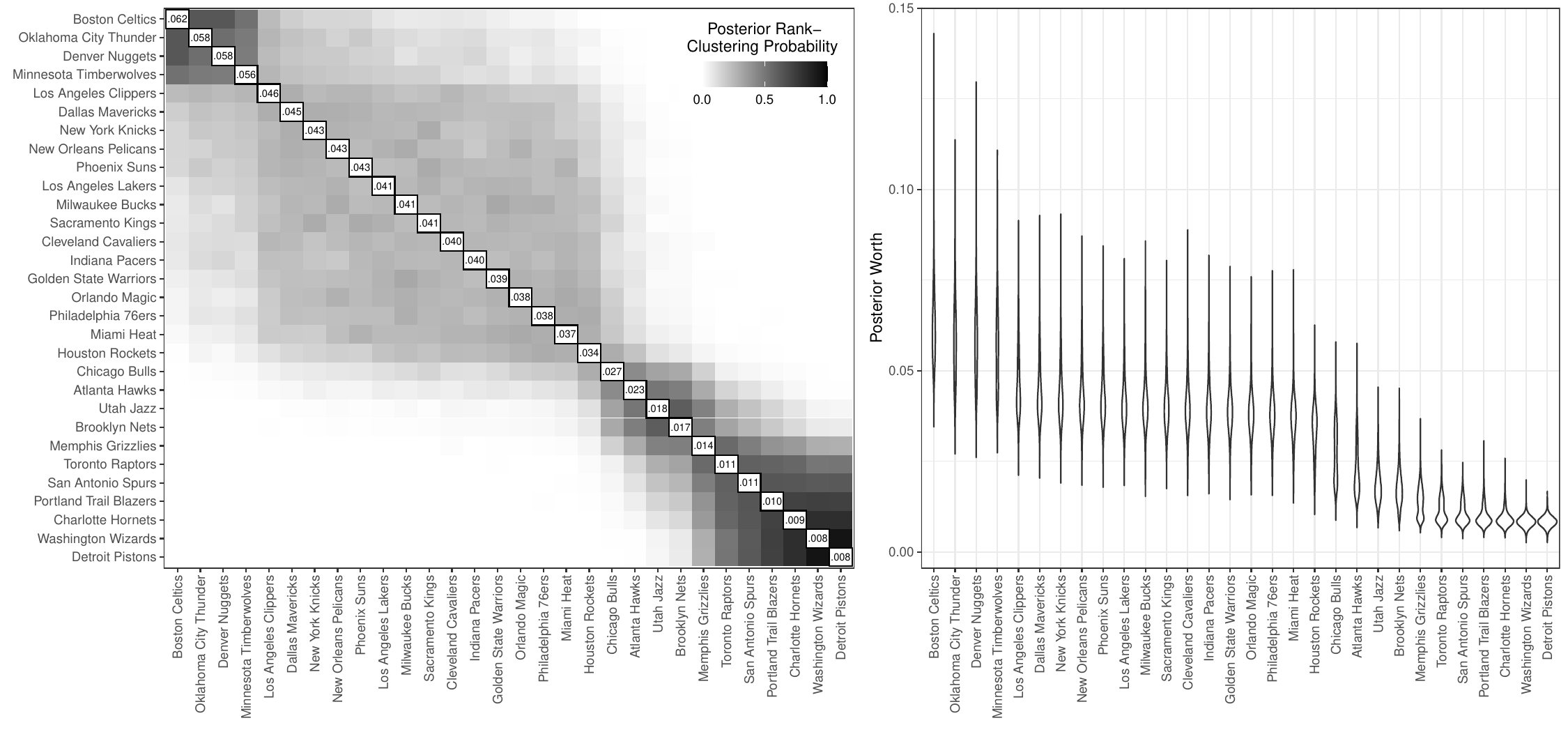}
    \caption{Primary results from Rank-Clustered BTL analysis of 2023-24 NBA data. \textit{Left:} Posterior rank-clustering probabilities. Main diagonal displays posterior median estimate of worth parameter after normalization. \textit{Right:} Posterior distributions of team-specific worth parameters.}
    \label{fig:NBA_1}
\end{figure}

We fit a Rank-Clustered Bradley-Terry-Luce model to the data with $a_\gamma=5$, $b_\gamma=3$, and $\lambda=1$ to encourage rank-clustering given the limited ordinal comparison data provided by pairwise matchups. Figure \ref{fig:NBA_1} displays posterior rank-clustering probabilities (left) and parameter posteriors (right). Additional results and convergence diagnostics are provided in the supplementary materials.

In this setting, the Rank-Clustered BTL model estimates an ordering of professional basketball teams with uncertain rank-clustering patterns. 
Uncertain rank-clustering may result from two aspects of this application. First, pairwise comparisons provide little information in relation to partial or complete rankings, by construction. Second, game outcomes provide low signal measurements of team ability \citep{baumer2023big}. That is because many factors influence game outcomes, such as skill, home advantage, injuries, roster changes, and luck \citep{cai2019hybrid}. Consistent with the low signal and limited information setting, an 80\% posterior credible interval indicates that there are between 6 and 9 rank-clusters. Every team has less than 0.037 posterior probability of belonging to a singleton rank-cluster. 

As seen in the left panel of Figure \ref{fig:NBA_1}, 4 teams (Boston Celtics, Oklahoma City Thunder, Denver Nuggets, and Minnesota Timberwolves) appear to be rank-clustered for first place. Based on regular season data alone, our model suggests that these 4 teams were of roughly indistinguishable ability.
Conversely, we observe that 6 teams (Toronto Raptors, San Antonio Spurs, Portland Trail Blazers, Charlotte Hornets, Washington Wizards, and Detroit Pistons) all have a high posterior probability of rank-clustering in last place. Instead of reporting the uncertain ranking of these teams with some granularity, we recommend to infer that these teams were the worst teams of the league in this season. These rank-clusters, despite not accounting for the complexities of the sport, provide useful and interpretable summaries of the teams' abilities across the regular season. A similar analysis could be used in the future to predict postseason performance.

\section{Discussion}\label{discussion}

In this paper, we proposed the Rank-Clustered Bradley-Terry-Luce model for estimating an overall ranking of objects with rank-clusters. The model employs the Bradley-Terry-Luce (BTL) family of distributions for ordinal comparisons. We proposed Partition-based Spike-and-Slab Fusion (PSSF) prior to estimates model parameters in a Bayesian framework. The model requires neither pre-specification of the number or size of the rank-clusters (improving upon \cite{piancastelli2024clustered}), nor specification of lasso-based penalty parameters (improving upon \cite{masarotto2012ranking,jeon2018sparse,hermes2024joint}). In a simulation study, we demonstrated the model's ability to accurately and consistently estimate the relative worth of objects in a collection while simultaneously estimating rank-clusters. We used Rank-Clustered BTL on four real datasets under different types of ordinal comparison data.


In contrast to the only other spike-and-slab based prior for parameter fusion~\cite{wu2021variable}, PSSF prior we developed does not require a known parameter order.
Visual inspection of the prior distribution makes obvious its connection to spike-and-slab: ``spike'' components correspond to parameter clusters and ``slab'' components correspond to independent parameters. Estimation of parameters under this model requires reversible jump MCMC. To overcome potentially slow or computationally-burdensome estimation in this setting, we proposed a computationally efficient Gibbs sampler. The sampler alternates between updating the partition of objects, based on the seminal work of \cite{green1995reversible}, and updating object-level worth parameters following a data augmentation trick for standard Plackett-Luce models by \cite{caron2012efficient} that was later adapted for Plackett-Luce mixtures by \cite{mollica2017bayesian}.

The proposed PSSF prior requires selecting hyperpriors for partitions, $f_G$, and the continuous values for each unique parameter, $f_\nu$. In this work, we specified $f_G\propto\text{Poisson}(K_g|\lambda)$ to be intentionally vague over the large space of partitions and $f_\nu\propto\text{Gamma}(a_\gamma,b_\gamma)$ based on conjugacy. However, alternative hyperpriors are available. A Negative Binomial or Beta Negative Binomial distribution for $f_G$ may be more appropriate when stronger prior knowledge of $K_g$ is available. If PSSF were to be applied to linear regression for parameter fusion, a Normal or $t-$distribution may be substituted for $f_\nu$.

A useful benefit of estimating parameter values and clusters in a single Bayesian framework is the avoidance of issues associated with \textit{selective inference} \citep{taylor2015statistical} or, more colloquially, \textit{double dipping} \citep{kriegeskorte2009circular}. Selective inference occurs when the same data is used twice in the process of model selection and/or estimation, e.g., to estimate some latent structure underlying the data and subsequently to estimate parameters conditional on that estimated structure. In our context, selective inference would occur if ordinal preference data was used first to identify rank-clusters and then used again to estimate worth parameter values conditional on those clusters. We note that selective inference occurs in the estimation of the related Clustered Mallows Model by \cite{piancastelli2024clustered}, which requires selecting the number and size of rank-clusters among objects before fitting the model. Selective inference often leads to invalid inference in part because uncertainty regarding the estimated clustering structure is not taken into account. However, Rank-Clustered Bradley-Terry-Luce models do not perform selective inference because parameter values and rank-clusters are estimated simultaneously. As such, our parameter estimates incorporate uncertainty across the posterior distributions of both the rank-clustering structure and the specific parameter values.

Results from Rank-Clustered BTL models are useful in a variety of inferential contexts. As noted in other fusion literatures on rankings, estimated overall rankings may be easier to understand and interpret when rank-clusters of objects are identified, as rank-clusters lead to fewer rank levels of objects to distinguish \citep{masarotto2012ranking}. In contexts where model results are used for prediction, such as in sports, estimating rank-clusters may improve predictive accuracy \citep{tutz2015extended}. Similarly, estimating rank-clusters is important in the context of decision-making: In peer review, for example, rank-clusters can be beneficial for communicating uncertainty in the assessment of preferences and for better transparency in funding decisions. We might imagine a scenario where a government agency is only able to fund two grants, however, two grant proposals are rank-clustered in second place. In this case, rank-clustering can be used to communicate uncertainty in the relative quality of the top proposals. A potential danger is that under this uncertainty, decision makers may be tempted to resort to unfair tie-breaking methods, e.g., selecting the proposal with the most famous author. Instead, tie-breaking should occur based on a fairer or more principled method, such as a partial lottery \citep{fang2016research, roumbanis2019peer, heyard2022rethinking}.

We list a few possible directions for future research. First, in this work we have not considered the level of interconnectedness among the assessed objects (e.g., if separate groups of judges assess completely distinct sets of objects). This is particularly relevant in the case of pairwise comparison data, in which some pairs of objects may never experience a head-to-head match-up. Second, the PSSF prior could be imposed as a prior for more complex BTL models or to other models entirely. In the former, the PSSF prior could be applied to preference learning via BTL distributions that incorporate covariates (e.g., \cite{baldassarre2023bradley,gormley2010clustering, chapaaan1982exploiting,hermes2024joint}). In that case, the prior may be modified to permit covariate parameter estimation in addition to rank-clustering. In the latter case, the PSSF prior may be applied to regression for variable fusion, and its performance may be compared to other existing Bayesian variable fusion methods (e.g.,  \cite{casella2010penalized,song2020bayesian,shimamura2019bayesian}). Third, we notice that the PSSF prior bears some resemblance to a Dirichlet process prior \citep{escobar1995bayesian}. The connection between Bayesian nonparametrics and Bayesian parameter fusion requires further study.

The proposed Rank-Clustered BTL model accurately estimates rank-clusters, permitting complex summaries beyond the traditional overall ranking and allowing for improved interpretability of the results. The Bayesian Rank-Clustered BTL model relies on a novel, spike-and-slab type prior for parameter fusion, and is estimated in a computationally-efficient manner. The applications in survey data, voting, and sports to aid informed inference and decision-making illustrate methodological versatility and broad applicability of our proposed rank-clustering approach.

\subsection*{Implementation}
Code to implement the Rank-Clustered BTL model may be found at \url{https://github.com/pearce790/rankclust}.

\subsection*{Acknowledgements}
The authors would like to thank Drs. T. Brendan Murphy and I. Claire Gormley for inspiring conversations on rank data modeling during the early stages of this work.

\subsection*{Funding}
The authors were supported by the National Science Foundation under Grant No. 2019901.


\bibliographystyle{chicago}
\bibliography{refs}

\newpage
\appendix
\gdef\thesection{Appendix \Alph{section}}

\section{Additional Application Results}

\subsection{Additional Results from Section 5.1}

\begin{figure}[ht]
    \centering
    \includegraphics[width=\textwidth]{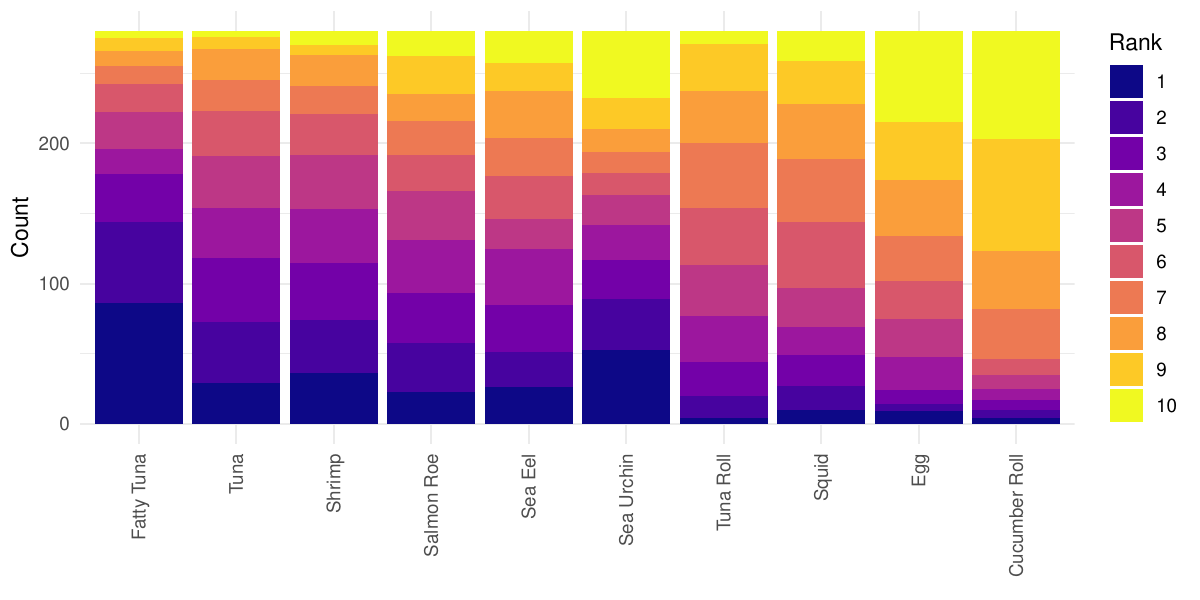}
   \caption{Stacked bar charts of ranks received by each sushi type.}
    \label{fig:sushi_eda}
\end{figure}

Figure \ref{fig:sushi_eda} displays stacked bar charts of ranks received by each sushi type by the survey respondents from Tohoku.

\begin{figure}[ht!]
    \centering
    \includegraphics[width=\textwidth]{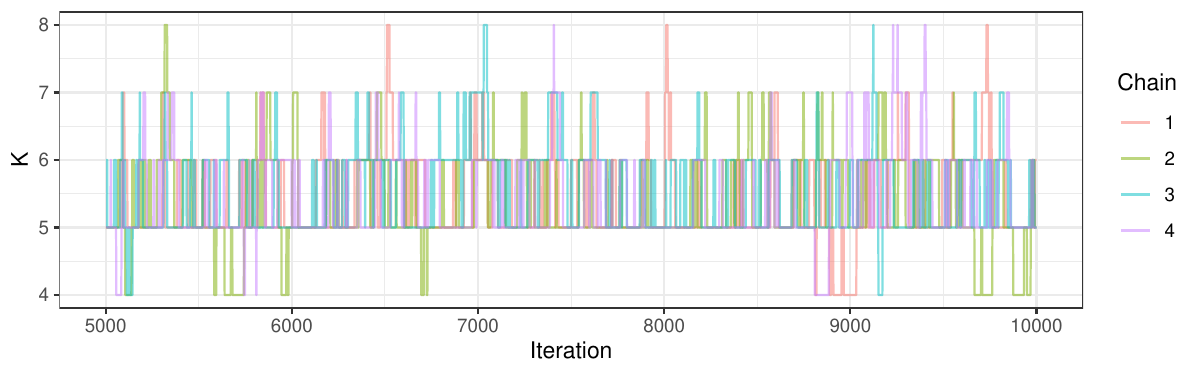}
    \caption{Trace plot of K in the Sushi data analysis}
    \label{fig:Sushi_2}
\end{figure}

\begin{figure}[ht!]
    \centering
    \includegraphics[width=\textwidth]{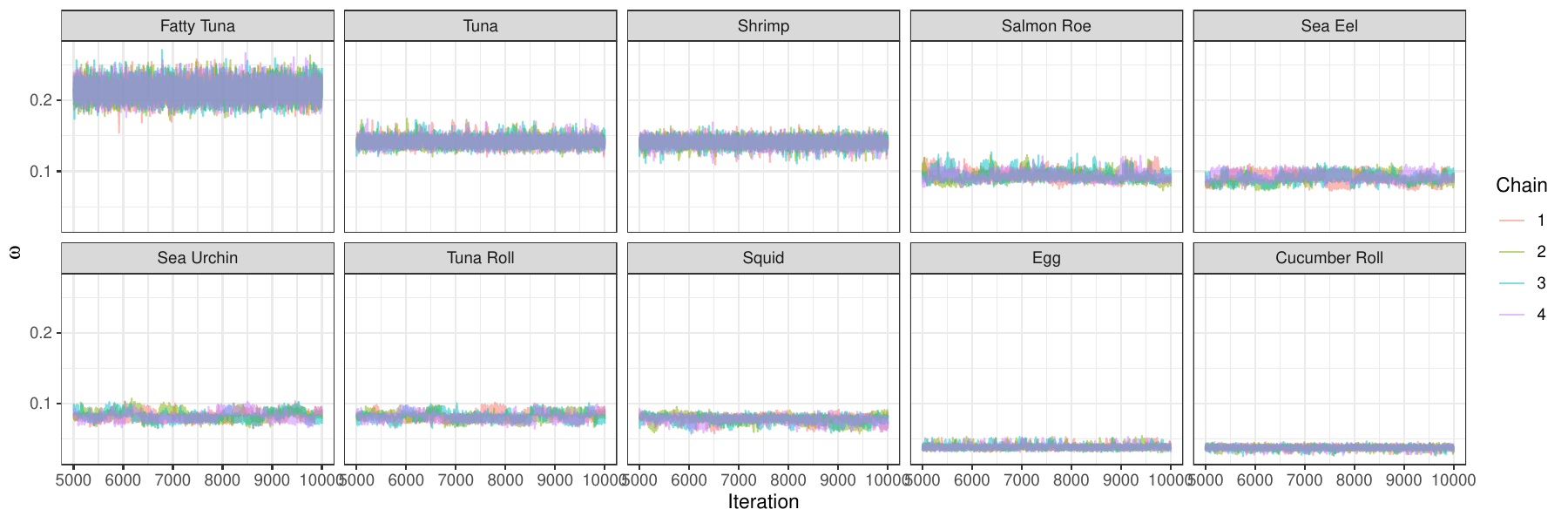}
    \caption{Trace plots of $\omega$ in the Sushi data analysis}
    \label{fig:Sushi_3}
\end{figure}

Figures \ref{fig:Sushi_2} and \ref{fig:Sushi_3} contain trace plots for $K$ and $\omega$ after burn-in for each chain. We find the trace plots to demonstrate satisfactory mixing and convergence.


\FloatBarrier

\subsection{Additional Results from Section 5.2}

\begin{figure}[ht]
    \centering
    \includegraphics[width=\textwidth]{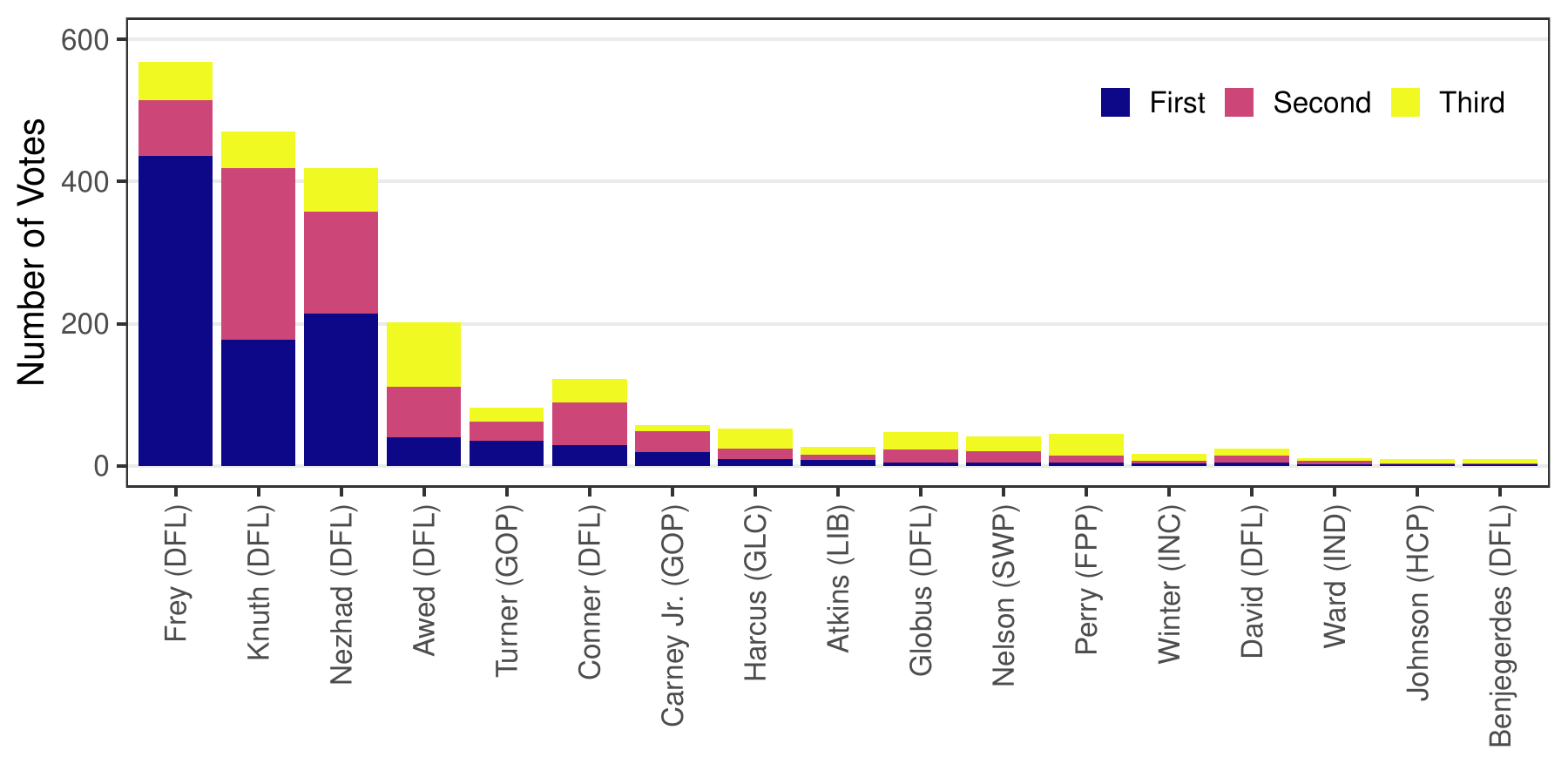}
   \caption[Exploratory analyses of 2021 Minneapolis mayoral election ranked choice voting data.]{Number of votes by rank level and candidate. Candidates are ordered by their position in the official ranked choice election. Acronyms on the tops of bars represent each candidate's political party.}
    \label{fig:votes_eda}
\end{figure}

Figure \ref{fig:votes_eda} displays stacked bar charts of the sampled votes by rank level for each candidate. 
Candidates are ordered by their final placement according to the official ranked choice voting algorithm. The incumbent, Jacob Frey, receives the largest share of first place votes, although Kate Knuth and Sheila Nezhad also receive substantial support. The remaining candidates receive comparatively few votes. Most candidates are associated with the Democratic-Farmer-Labor (DFL) party, which is affiliated with the national Democratic Party. Laverne Turner and Bob ``Again'' Carney Jr. are the only Republicans (GOP) in the race. The remaining candidates represent Grassroots--Legalize Cannabis (GLC), Libertarian (LIB), Socialist Workers Party (SWP), For the People Party (FPP), Independence (INC), Independent (IND), and Humanitarian--Community Party (HCP).

\begin{figure}[ht]
    \centering
    \includegraphics[width=\textwidth]{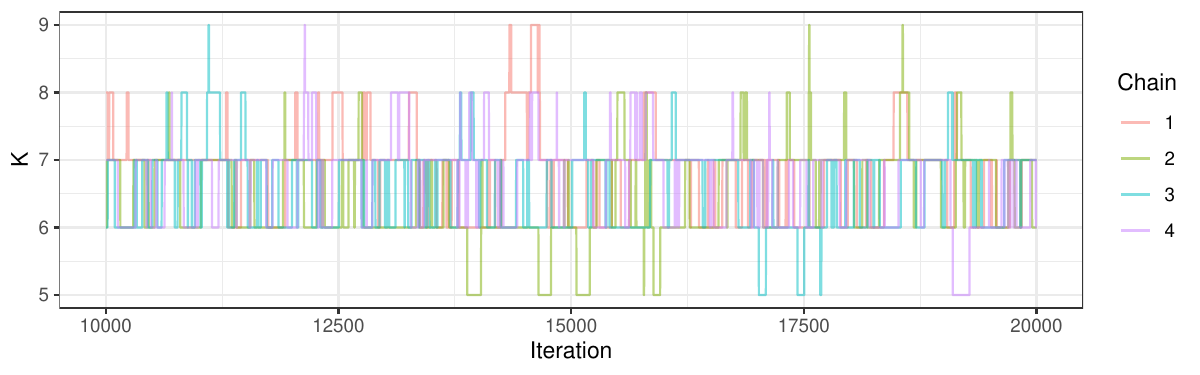}
    \caption{Trace plots of $K$ in the Minneapolis mayoral election data analysis.}
    \label{fig:objectties:K_trace}
\end{figure}

\begin{figure}[ht]
    \centering
    \includegraphics[width=.95\textwidth]{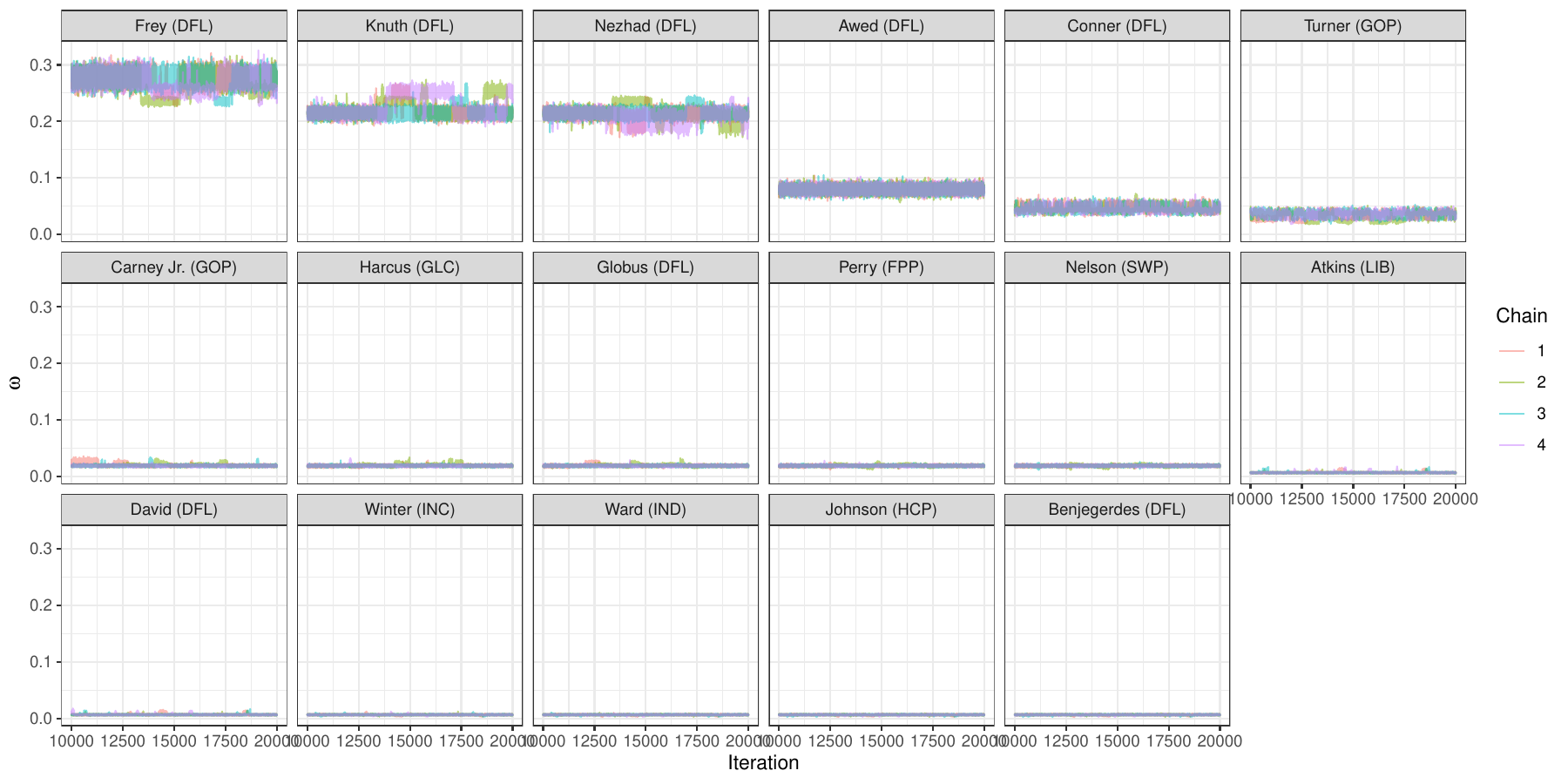}
    \caption{Trace plots of $\omega$ in the Minneapolis mayoral election data analysis.}
    \label{fig:objectties:Omega_trace}
\end{figure}

Figures \ref{fig:objectties:K_trace} and \ref{fig:objectties:Omega_trace} contain trace plots for $K$ and $\omega$ after burn-in for each chain. We find the trace plots to demonstrate satisfactory mixing and convergence.

\FloatBarrier

\subsection{Additional Results from Section 5.3}

\begin{figure}[ht]
    \centering
    \includegraphics[width=\textwidth]{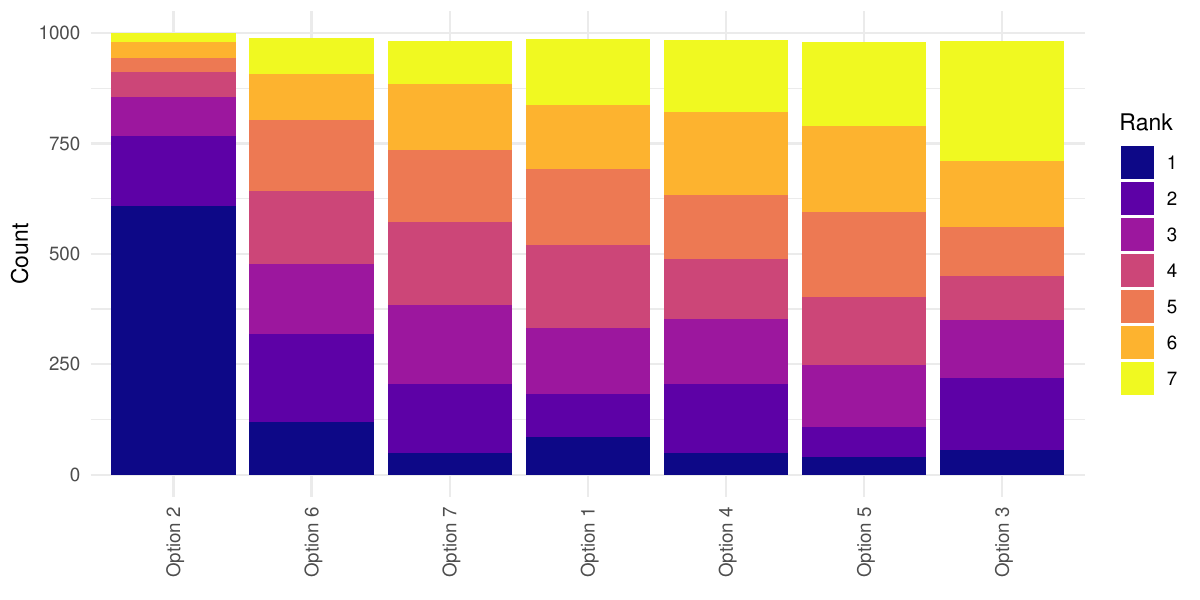}
   \caption{Stacked bar charts of ranks received by each policy option.}
    \label{fig:euro_eda}
\end{figure}

Figure \ref{fig:euro_eda} displays stacked bar charts of ranks received by each policy option by the survey respondents from Great Britain.

\begin{figure}[ht!]
    \centering
    \includegraphics[width=\textwidth]{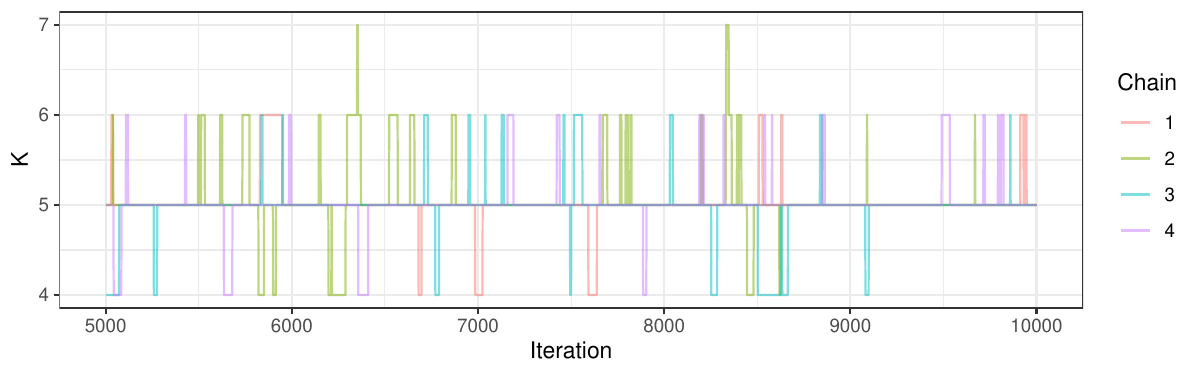}
    \caption{Trace plot of K in the Eurobarometer survey data analysis}
    \label{fig:EB_2}
\end{figure}

\begin{figure}[ht!]
    \centering
    \includegraphics[width=\textwidth]{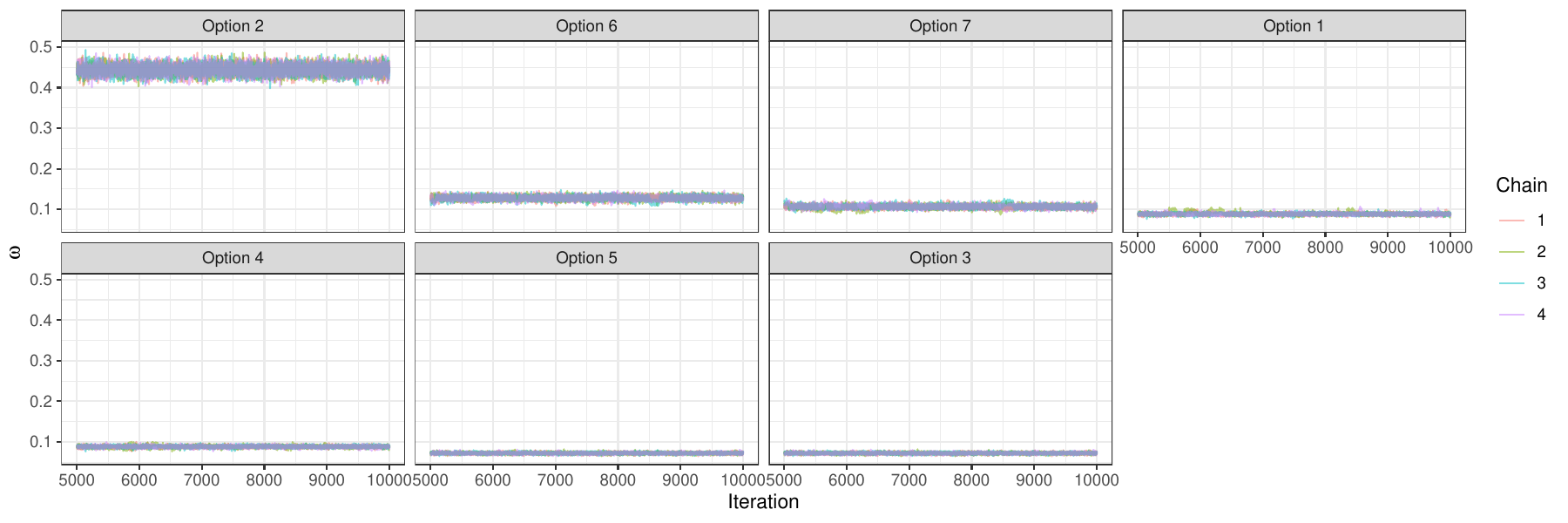}
    \caption{Trace plot of $\omega$ in the Eurobarometer survey data analysis}
    \label{fig:EB_3}
\end{figure}

Figures \ref{fig:EB_2} and \ref{fig:EB_3} contain trace plots for $K$ and $\omega$ after burn-in for each chain. We find the trace plots to demonstrate satisfactory mixing and convergence.


\FloatBarrier

\subsection{Additional Results from Section 5.4}

\begin{figure}[ht]
    \centering
    \includegraphics[width=.9\textwidth]{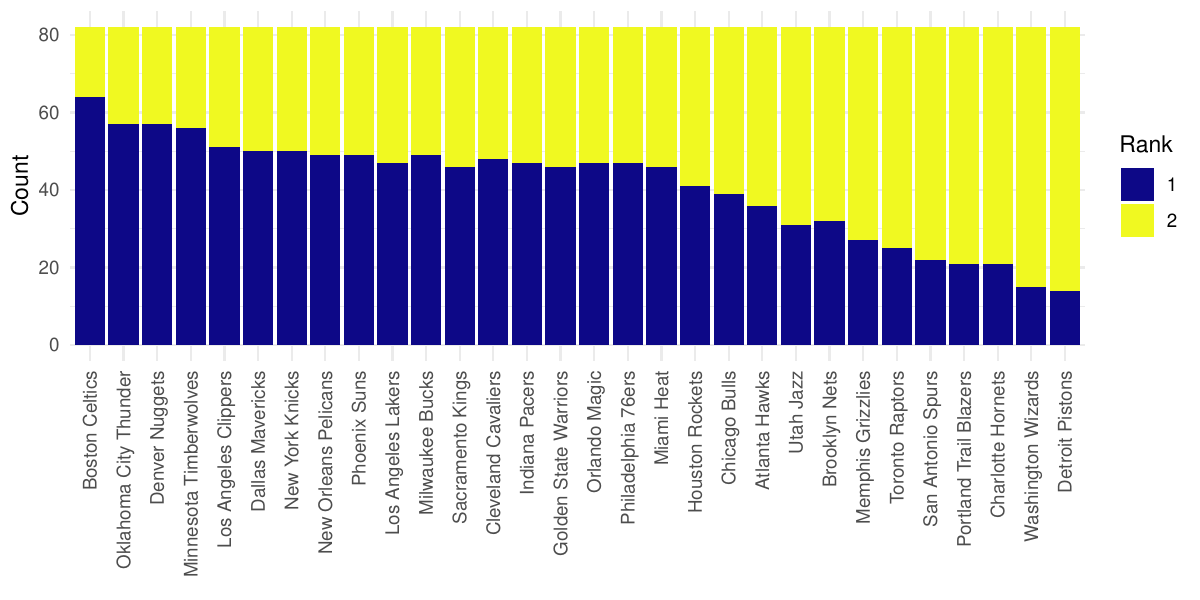}
   \caption{Stacked bar charts of ranks received by each NBA team across the 2023-24 season. Winning = rank 1; losing = rank 2.}
    \label{fig:NBA_eda}
\end{figure}

Figure \ref{fig:NBA_eda} displays stacked bar charts of the season record of each NBA team across the 2023-24 season.

\begin{figure}[ht!]
    \centering
    \includegraphics[width=\textwidth]{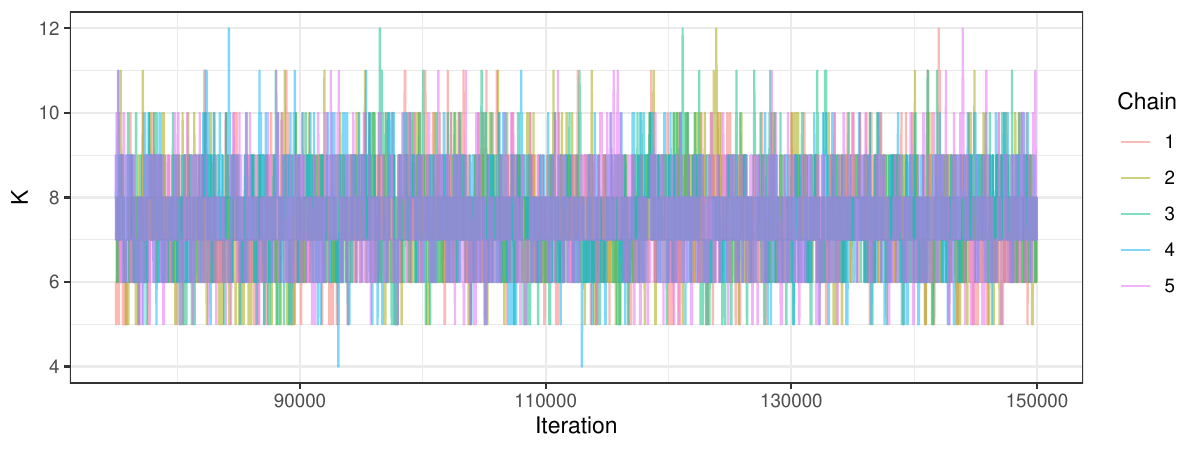}
    \caption{Trace plot of K in the 2023-24 NBA season analysis}
    \label{fig:NBA_2}
\end{figure}

\begin{figure}[ht!]
    \centering
    \includegraphics[width=\textwidth]{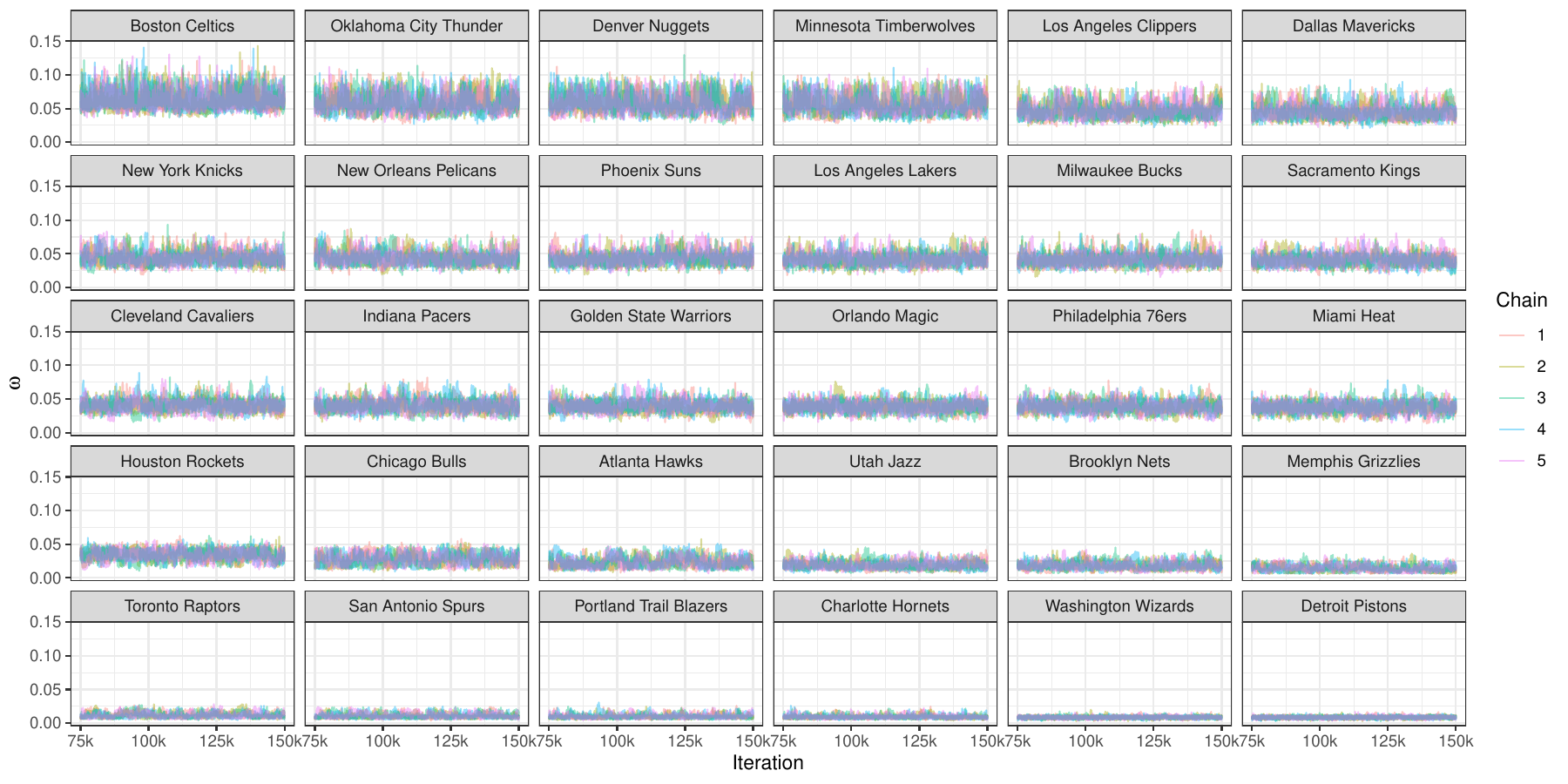}
    \caption{Trace plot of $\omega$ in the 2023-24 NBA season analysis}
    \label{fig:NBA_3}
\end{figure}

Figures \ref{fig:NBA_2} and \ref{fig:NBA_3} contain trace plots for $K$ and $\omega$ after burn-in for each chain. We find the trace plots to demonstrate satisfactory mixing and convergence.


\end{document}